\documentclass[fleqn,10pt,twocolumn]{wlscirep}
\usepackage[utf8]{inputenc}
\usepackage[T1]{fontenc}
\usepackage{color}
\usepackage{ragged2e}
\usepackage{dcolumn}
\usepackage{amsmath}
\usepackage{amssymb}
\usepackage{epsfig}
\usepackage{float}
\usepackage{graphicx}
\usepackage{subfigure}
\usepackage{hyperref}
\usepackage{color}
\graphicspath{{images/}}

\begin{document}

\title{Integrated distributed sensing and quantum communication networks}

\author[1]{Yuehan Xu}
\author[1,2,3,*]{Tao Wang}
\author[1,2,3]{Peng Huang}
\author[1,2,3,$\dag$]{Guihua Zeng}

\affil[1]{State Key Laboratory of Advanced Optical Communication Systems and Networks, Center of Quantum Sensing and Information Processing, Shanghai Jiao Tong University, Shanghai 200240, China}
\affil[2]{Shanghai Research Center for Quantum Sciences, Shanghai 201315, China}
\affil[3]{Hefei National Laboratory, Hefei 230088, China}
\affil[*]{tonystar@sjtu.edu.cn}
\affil[$\dag$]{ghzeng@sjtu.edu.cn}

\begin{abstract}

The integration of sensing and communication can achieve ubiquitous sensing while enabling ubiquitous communication. Within the gradually improving global communication, the integrated sensing and communication (ISAC) system based on optical fibers can accomplish various functionalities, such as urban structure imaging, seismic wave detection, and pipeline safety monitoring. With the development of quantum communication, quantum networks based on optical fiber are gradually being established. In this paper, we propose an integrated sensing and quantum network (ISAQN) scheme, which can achieve secure key distribution among multiple nodes and distributed sensing under the standard quantum limit. The continuous variables quantum key distribution (CV-QKD) protocol and the round-trip multi-band structure are adopted to achieve the multi-node secure key distribution. Meanwhile, the spectrum phase monitoring (SPM) protocol is proposed to realize distributed sensing. It determines which node is vibrating by monitoring the frequency spectrum and restores the vibration waveform by monitoring the phase change. The scheme is experimentally demonstrated by simulating the vibration in a star structure network. Experimental results indicate that this multi-user quantum network can achieve a secret key rate (SKR) of approximately 0.7 $\rm{Mbits/s}$ for each user under 10 $\rm{km}$ standard fiber transmission and its network capacity is 8. In terms of distributed sensing, it can achieve a vibration response bandwidth ranging from 1 $\rm{Hz}$ to 2 $\rm{kHz}$, a strain resolution of 0.50 $\rm{n}$$\varepsilon$$/\sqrt{\rm{Hz}}$, and a spatial resolution of 0.20 $\rm{m}$ under shot-noise-limited detection. The proposed ISAQN scheme enables simultaneous quantum communication and distributed sensing in a multi-point network, laying a foundation for future large-scale quantum networks and high-precision sensing networks.

\end{abstract}

\flushbottom
\maketitle

\thispagestyle{empty}

\section*{Introduction}

In recent years, various communication networks have been applied worldwide, creating an era of interconnected information. Due to the sensitivity of communication media to environmental changes, communication media can be used not only for information transmission but also for sensing. Thus, the concept of integrated sensing and communication (ISAC) emerged. ISAC requires sensing to be conducted simultaneously during the communication process, rather than building a separate sensing network. Its goal is to achieve ubiquitous integrated sensing and communication networks, providing large-scale sensing for urban structure imaging, seismic wave detection, pipeline safety monitoring, etc. Initially, ISAC was proposed in wireless communication. Recently, ISAC in optical communication has been implemented, enabling sensing demodulation while conducting optical communication. By using optical fibers as the transmission medium, ISAC can achieve high-speed communication while also achieving high-precision sensing.

Quantum communication is one of the most concerned communication methods. Quantum key distribution (QKD) is the core technology of quantum communication, providing secure keys for legitimate parties guaranteed by the basic principles of quantum mechanics \cite{gisin2002quantum}. QKD was proposed in 1984, and researchers have currently completed its security proof, experimental verification, field verification, chip integration, and prototype implementation \cite{grosshans2002continuous,grosshans2003quantum,weedbrook2004quantum,renner2009finetti,leverrier2009unconditional,leverrier2013security,jouguet2013experimental,qi2015generating,soh2015self,huang2016long,kleis2017continuous,leverrier2017security,zhang2019integrated,zhang2020long,wang2020high,ren2021demonstration,xu2023simultaneous}. Beyond traditional point-to-point QKD system, QKD network \cite{dianati2008architecture,stucki2011long,wang2014field,bedington2017progress,tajima2017quantum,kiktenko2017demonstration,zhang2018large,xu2023round} can ensure that multiple users share quantum secure keys. It can be classified into backbone networks, metropolitan area networks, and access networks based on their coverage area. A typical backbone QKD network is the Beijing-Shanghai trunk line \cite{chen2021integrated}, which achieves QKD transmission at a distance of over 4600 $\rm{km}$. In addition, Cambridge QKD metropolitan area network is constructed with high bandwidth data transmission \cite{dynes2019cambridge}, which has been operating for several years with 3 nodes separated by 5-10 $\rm{km}$ optical fiber \cite{wonfor2021quantum}. The 46-node QKD metropolitan area network in Hefei realizes real-time voice telephone, text messaging, and file transmission \cite{chen2021implementation}. The implementation of a QKD access network for multiple users was proposed by Bernd Frohlich et al. through an upstream quantum access network \cite{frohlich2013quantum}.

In the practical implementation of QKD, optical signals are used as the carrier for transmitting secure key information. They are very sensitive to changes in phase, amplitude, wavelength, etc., and environmental disturbances can cause corresponding changes. Pilot signals are commonly employed to eliminate those effects on QKD signals. This characteristic can in turn be used for distributed optical fiber sensing (DOFS). DOFS technology utilizes specific effects of vibration, acoustic, and temperature on the phase, amplitude, and wavelength of light in the optical fiber to achieve distributed fiber vibration sensing (DVS), distributed fiber acoustic sensing (DAS), and distributed fiber temperature sensing (DTS) \cite{taylor1993apparatus,pan2011phase}. It is mainly divided into two types. One uses back-scattering light to achieve sensing also known as phase-sensitive optical time-domain reflectometry ($\phi$-OTDR\cite{juarez2005distributed,selker2006distributed,zhang2008distributed,koyamada2009fiber,lu2010distributed,peng2014ultra,dong2016quantitative,chen2017distributed,chen2018high,lu2019distributed}), while the other uses the forward-transmitting light to achieve sensing. The back-scattering light sensing scheme can achieve a spatial resolution of 0.8 $\rm{m}$, a strain resolution of about 0.25 $\rm{n}$$\varepsilon$$/\sqrt{\rm{Hz}}$, and a vibration response bandwidth of  5 $\rm{kHz}$ along the total 9.8 $\rm{km}$ sensing fiber \cite{chen2018high}. The forward-transmitting light scheme is capable of detecting earthquakes over terrestrial and submarine links with lengths ranging from 75 to 535 $\rm{km}$ and a geographical distance from the earthquake’s epicenter ranging from 25 to 18,500 $\rm{km}$\cite{marra2018ultrastable}. In addition, it can achieve a spatial resolution of 1 $\rm{km}$ in the frequency calibration fiber link of QKD \cite{chen2022quantum}.

Recently, DOFS has been integrated into classical optical communication to enable both communication and sensing, thereby constructing an ISAC system \cite{huang2019first,huang2020simultaneous,guerrier2022vibration,ip2022over,he2023integrated}. Correspondingly, the integrated sensing and quantum communication (ISAQC) system requires simultaneous sensing and QKD without additional devices, while ensuring there is no interference between the two procedures. Furthermore, an integrated sensing and quantum network (ISAQN) composed of multiple ISAQC systems requires the differentiation of QKD signals and sensing signals from different nodes. Each network node can serve as a QKD node and also serve as a sensing node, thereby achieving ISAQN. However, two issues urgently need to be addressed. The first lies in the inability of weak QKD signals to achieve precise sensing, thus making it impossible to perform high-precision sensing based on QKD signals. Specifically, the average number of photons in QKD signals is very small, whereas the DOFS based on back-scattering requires a strong optical signal. The second lies in the difficulty of discerning between multiple QKD signals and sensing signals. Since both sensing information and key information are loaded onto the same coherent state, extracting the key information and demodulating the sensing information pose challenges. Therefore, a reasonable scheme is needed to implement ISAQN.

In order to achieve signal sensing in multi-point QKD networks, we propose the time-frequency-multiplexing ISAQN. This scheme utilizes both quantum signals and pilot signals for transmission. Quantum signals are used for continuous variables QKD (CV-QKD), while pilot signals are used for sensing. Additionally, the pilot signal inserted by the time division multiplexing (TDM) can restore the phase of the quantum signal when there is no vibration. For multiple nodes, we use frequency division multiplexing (FDM) to transmit multiple QKD signals of different nodes. When the vibration is happening, the nodes experiencing vibrations can be determined and the vibration waveform can be restored based on the spectrum phase monitoring (SPM) protocol. The precision of ISAQN's distributed sensing can reach the standard quantum limit. To demonstrate the feasibility of our scheme, experiments were carried out by simulating the vibration in a star network structure with a network capacity of 8. The results indicate that this network can achieve a secret key rate (SKR) of approximately 0.7 $\rm{Mbits/s}$ under 10 $\rm{km}$ standard fiber transmission, a vibration response bandwidth ranging from 1 $\rm{Hz}$ to 2 $\rm{kHz}$, a strain resolution of 0.50 $\rm{n}$$\varepsilon$$/\sqrt{\rm{Hz}}$, and a spatial resolution of 0.20 $\rm{m}$ under shot-noise-limited detection.

In this paper, our ISAQN scheme is introduced in detail. First, we describe the physical structure of ISAQN and how it works. In addition, we have analyzed the theoretical principles of CV-QKD and DOFS. Based on this physical structure, we construct a proof-of-principle experiment and verify the feasibility of the ISAQN. Finally, we come up with a conclusion.
 
\section*{Result}

\subsection*{Integrated distributed sensing and quantum communication networks}

In order to achieve point-to-multipoint quantum communication and distributed sensing in the same network, we propose the time-frequency-multiplexing ISAQN. This network conducts quantum communication and sensing simultaneously through coherent states. Firstly, we will describe this point-to-multipoint quantum network in part a. Secondly, we will explain how distributed sensing is integrated into this quantum network in part b.

\subsubsection*{a. Point-to-multipoint quantum networks}

\begin{figure}[htb]
\centering
\includegraphics[width=1\linewidth]{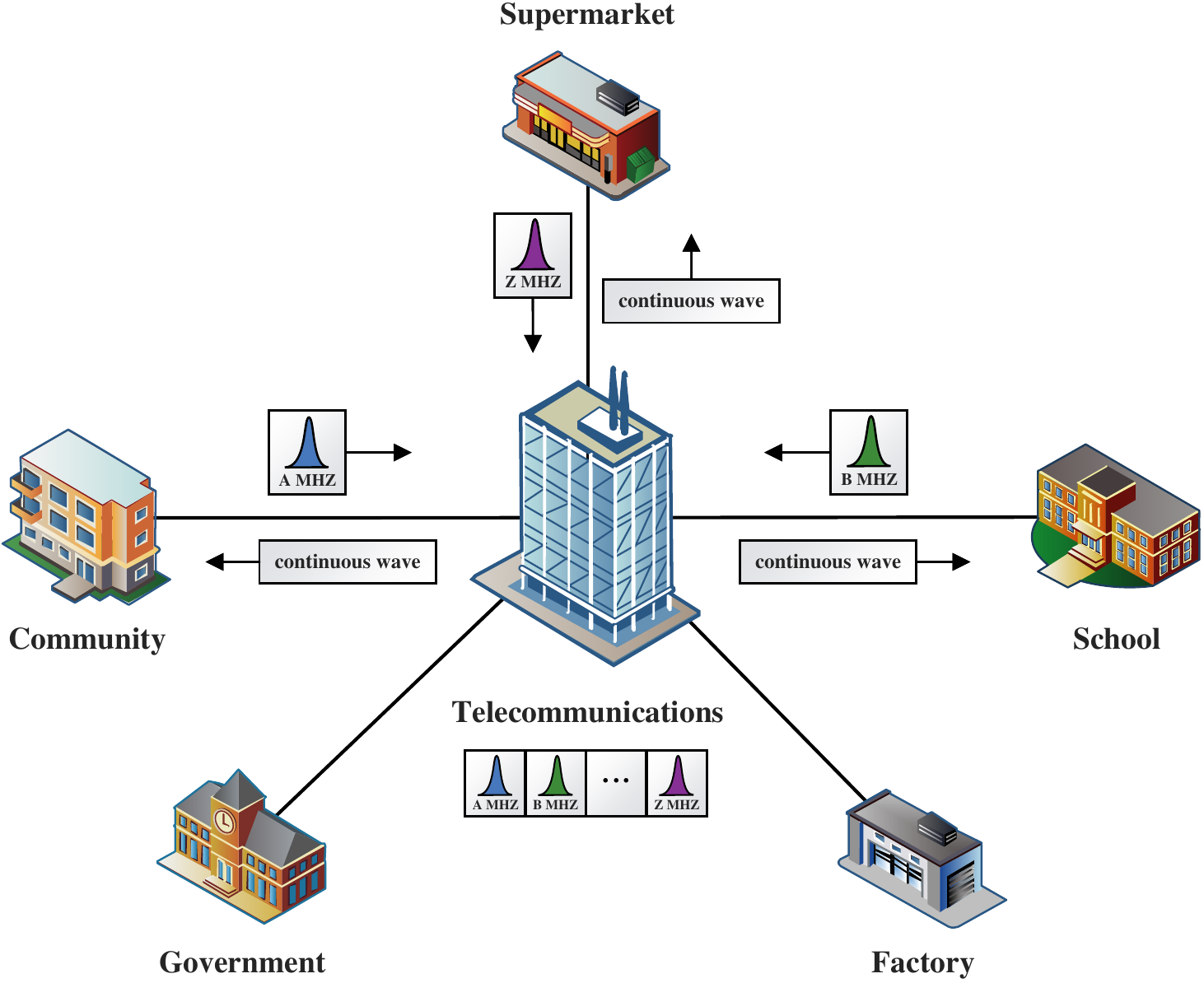}
\caption{Schematic diagram of the round-trip multi-band quantum key distribution (QKD) network. The Telecommunication as center node performs point-to-multipoint QKD with Community, Supermarket, School, Government, and Factory as child nodes. A continuous wave is transmitted from the center node to each child node. Every child node modulates quantum signals at their respective carrier frequencies and transmits them back to the center node. Subsequently, these signals are superimposed to form a multi-band signal at the center node.}
\label{fig1}
\end{figure}

\begin{table}[!h]
	\centering
	\renewcommand\arraystretch{1.2}
	\begin{tabular}{c p{7cm}}
		\hline
		\hline
		\textbf{1.} & Alice selects two sets of Gaussian distributed random sequences $x_A$ and $p_A$, with length $n$, mean 0, and variance of $V_A$. Based on these, $n$ coherent states $\left|x_{A}+jp_{A} \right\rangle$ are prepared by Alice, and then transmitted to Bob through a quantum channel. The quantum channel can be characterized by the transmittance $T$ and the noise $\varepsilon$. \\
		\hline	
		\textbf{2.} & After receiving the coherent state, Bob will simultaneously measure both quadrature components, which is heterodyne detection. The measurement results are denoted as $x_B$ and $p_B$. The practical detector at Bob's input can be characterized with the quantum efficiency $\eta$ and the electrical noise $v_{el}$. \\
		\hline
		\textbf{3.} & Because heterodyne detection is adopted, Alice retains all the data. \\
		\hline
		\textbf{4.} & Alice randomly selects a portion of the retained data for parameter evaluation and publicly discloses this data. Bob estimates parameters based on the measurement data, including channel transmittance, channel excess noise, and modulation variance. Then, Bob evaluates the secure key rate using these parameters. If the secure key rate is less than 0, the key distribution is terminated and retransmitted. \\
	    \hline
	    \textbf{5.} & Alice and Bob perform data post-processing on the remaining data, including steps such as reverse reconciliation and privacy amplification. Eventually, both parties obtain the same secure key of $m$ $\rm{bits}$. \\
		\hline
		\hline
	\end{tabular}
	\caption{The steps of point-to-point Gaussian modulated coherent states (GMCS) continuous variables QKD (CV-QKD)}
	\label{tab1}
\end{table}

 CV-QKD uses coherent states to distribute secure keys, thus ensuring the security of communication. The commonly used point-to-point Gaussian modulated coherent states (GMCS) CV-QKD protocol is shown in Tab. \ref{tab1}. However, this protocol does not directly support multi-point QKD. Therefore, we propose the round-trip structure for constructing the point-to-multipoint QKD network \cite{xu2023round}, shown in Fig. \ref{fig1}. This network enables QKD between the center node and each of the child nodes. It can be understood that the child node is Alice and the center node is Bob in the classical scheme. As illustrated in Fig. \ref{fig1}, the Telecommunication as center node performs point-to-multipoint QKD with Community, Supermarket, School, Government, and Factory as child nodes. In the round-trip structure, the center node holds the laser and detector for transmitting and receiving light, while each child node only performs quantum signal modulation. This process can be divided into two steps. First, light is transmitted from the center node to each child node, with every child node receiving $1/N$ of the light, where $N$ is the network capacity. Subsequently, each child node modulates the quantum signal with information and transmits it back to the center node, resulting in a $1/N$ transmittance loss. The modulated coherent states from each child node are received at the center node, which aggregates the quantum state transmitted from every node. Therefore, it only requires a laser and a detector to efficiently complete a point-to-multipoint QKD network.

For the multi-point QKD network, differentiating quantum signals from different nodes is also a challenge. In the practical implementation, we use FDM to differentiate individual child nodes in the point-to-multipoint QKD network. FDM is a multiplexing technique that facilitates the modulation of multiple base-band signals onto various carrier frequencies. In other words, different frequency bands are multiplexed by modulating the carrier wave at different rates, which can be equivalently represented as different rotation speeds in the phase space. In Fig. \ref{fig1}, when each child node receives a continuous wave, they modulate quantum signals at their respective carrier frequencies and transmit them back to the center node. Subsequently, these signals are superimposed to form a multi-band signal at the center node. During the demodulation process, different frequency bands are filtered to distinguish different child nodes. Hence, the multi-band structure brought by FDM technology is suitable for QKD in multi-user networks. Although TDM is another feasible alternative, it necessitates tighter control over time slots.

In the following, we will provide a detailed description of the specific principles for implementing the round-trip multi-band QKD network. First, the center node transmits a continuous wave to each child node. The $i$-th child node selects Gaussian distribution values $x_i$ and $p_i$ with equal probability, producing a random sequence of length $n$, mean 0, and variance of $V_i$. Then, the $i$-th child node creates $n$ coherent states based on this random sequence, which can be represented as
\begin{equation}
	\left|\alpha_{i}\right\rangle=\left|x_{i}+jp_{i} \right\rangle e^{j\omega_i t},i\in\{1,2,\cdots,N \},
\label{formula1}
\end{equation}
where $j$ is the imaginary number unit, $N$ denotes the network capacity, $\alpha_{i}^{2}=V_i/2=V_{\mathrm{A}}/2$, $\omega_i$ is the carrier frequency of the $i$-th child node, and $t$ represents time. We can expand the formula for the $i$-th child node's coherent state as
\begin{equation}
\begin{aligned}
	\left|x_{i}+jp_{i} \right\rangle e^{j\omega_i t} &=\left|\left(x_{i} \cos \left(\omega_i t \right)-p_{i}\sin \left(\omega_i t \right)\right) \right. \\
	&\left. +j\left(x_{i} \sin \left(\omega_i t \right)+p_{i} \cos \left(\omega_i t \right)\right) \right\rangle,
\label{formula2}
\end{aligned}
\end{equation}
where the real part $x_{i} \cos \left(\omega_i t \right)-p_{i} \sin \left(\omega_i t \right)$ and the imaginary part $x_{i} \sin \left(\omega_i t \right)+p_{i} \cos \left(\omega_i t \right)$ refer to the radio frequency (RF) signals respectively added to the two paths of the in-phase and quadrature modulator (IQM). When transmitted through optical fibers, the coherent state undergoes a rotation of angle $\theta$. Each child node experiences a different rotation angle, and $\theta_{i}$ represents the rotation angle of the $i$-th child node. The splitter collects the signals modulated by every child node to form a signal with $N$ frequency bands. Finally, the signal returns to the center node through the round-trip structure. Therefore, the coherent state received by the coherent detector can be expressed as
\begin{equation}
\begin{aligned}
	&\sum_{i=1}^{n} \left|\alpha_{i}\right\rangle e^{j \theta_{i}} \\
	&= \sum_{i=1}^{n} \left|x_{i}+jp_{i} \right\rangle e^{j\left(\omega_i t+\theta_{i}\right)} \\
	&= \sum_{i=1}^{n} \left|\left(x_{i}+jp_{i} \right) \cos(\omega_i t + \theta_{i}) + j\left(x_{i}+jp_{i} \right) \sin(\omega_i t + \theta_{i}) \right\rangle.
\end{aligned}
\end{equation}
The center node obtains a spectrum consisting of mixed multi-frequency bands, which are difficult to distinguish in the time domain but can be clearly distinguished in the frequency domain. Then, the center node checks the frequency bands registered by the child nodes. The center node examines the registered frequency bands to determine which child nodes are currently communicating. This registration method can effectively prevent the unauthorized use of illegal frequency bands by Eve to steal information. For these legal bands, the center node uses band-pass filtering to isolate them. Therefore, the quantum signal of the $i$-th child node can be separated and expressed as
\begin{equation}
\begin{aligned}
	\left|\alpha_{i}\right\rangle e^{j \theta_{i}} & = \left|x_{i}+jp_{i} \right\rangle e^{j \left(\omega_i t+\theta_{i}\right)} \\
	& = \left|\left(x_{i}+jp_{i} \right) \cos(\omega_i t + \theta_{i}) + j\left(x_{i}+jp_{i} \right) \sin(\omega_i t + \theta_{i}) \right\rangle \\
	& = \left|\left(x_{i}+jp_{i} \right) (\cos(\omega_i t)\cos\theta_{i} - \sin(\omega_i t)\sin\theta_{i}) \right. \\
	& \left. + j\left(x_{i}+jp_{i} \right) (\sin(\omega_i t)\cos\theta_{i} + \cos(\omega_i t)\sin\theta_{i}) \right\rangle \\
	& = \left|\left(x_{i} \cos \left(\omega_i t \right)\cos \theta_{i} - x_{i} \sin \left(\omega_i t \right)\sin \theta_{i} \right. \right. \\
	& \left. \left. -p_{i}\sin \left(\omega_i t \right)\cos \theta_{i} - p_{i} \cos\left(\omega_i t \right)\sin \theta_{i} \right) \right. \\
	& \left.+j\left(x_{i} \sin \left(\omega_i t \right)\cos \theta_{i} + x_{i} \cos \left(\omega_i t \right)\sin \theta_{i} \right. \right. \\
	& \left. \left. +p_{i} \cos \left(\omega_i t \right)\cos \theta_{i} - p_{i} \sin \left(\omega_i t \right)\sin \theta_{i} \right) \right \rangle,
\end{aligned}
\end{equation}
where the real part $\left|Re\left(\alpha_{i} e^{j \theta_{i}}\right)\right\rangle$ and imaginary part $\left|Im\left(\alpha_{i} e^{j \theta_{i}}\right)\right\rangle$ are respectively the two results detected by the heterodyne detection (also known as dual-homodyne detection). The center node can eliminate the carrier frequency of every child node via coherent demodulation, obtaining base-band signals. The specific steps are as follows. First, the real part and the imaginary part are multiplied by $\cos\left(\omega_i t \right)$ to obtain
\begin{equation}
\begin{aligned}
	&\left|Re\left(\alpha_{i} e^{j \theta_{i}}\right)\right\rangle \cos \left(\omega_i t \right) \\
	&= \left| \frac{1}{2} \left(\cos \left(2\omega_i t \right)+1\right)\left(x_{i} \cos \theta_{i} - p_{i} \sin \theta_{i}\right) \right. \\
	&\left. - \frac{1}{2}\sin \left(2\omega_i t \right)\left(x_{i} \sin \theta_{i}+ p_{i} \cos \theta_{i} \right) \right\rangle, \\
	&\left|Im\left(\alpha_{i} e^{j \theta_{i}}\right)\right\rangle \cos\left(\omega_i t \right) \\
	&= \left| \frac{1}{2} \left(\cos \left(2\omega_i t \right)+1\right)\left(x_{i} \sin \theta_{i} + p_{i} \cos \theta_{i}\right) \right. \\
	&\left. + \frac{1}{2}\sin \left(2\omega_i t \right)\left(x_{i} \cos \theta_{i} - p_{i} \sin \theta_{i} \right) \right\rangle.
\end{aligned}
\end{equation}
After filtering out the high-frequency components with a low-pass filter, we can obtain
\begin{equation}
\begin{aligned}
	\left|X_i\right\rangle=\left|x_{i} \cos \theta_{i} - p_{i} \sin \theta_{i} \right\rangle,\\
	\left|P_i\right\rangle=\left|x_{i} \sin \theta_{i} + p_{i} \cos \theta_{i} \right\rangle.
\end{aligned}
\end{equation}
Then, we use the pilot signal inserted in the quantum signal by the TDM method to estimate $\theta$, and then recover the quantum signal. The pilot signal corresponds to the state of $x_i=1$ and $p_i=0$. The angle $\theta_i$ can be acquired as
\begin{equation}
\begin{aligned}
	\theta_i =\arctan \left( \left| X_i \right\rangle / \left| P_i \right\rangle \right).
\end{aligned}
\end{equation}
Subsequently, the center node performs frame synchronization to align the data. Upon completion of parameter evaluation, reverse reconciliation, and privacy amplification, the QKD process is finalized. We summarized the process of point-to-multipoint GMCS CV-QKD in Tab. \ref{tab2}. In this scheme, every child node requires only one modulator and one circulator to access the round-trip multi-band QKD network, with only one laser and detector required for the entire network, which is highly efficient. This scheme avoids the issue of wavelength misalignment caused by different laser sources. The detector bandwidth is fully utilized due to the FDM method. Furthermore, the network possesses remarkable scalability, robustness, and noise suppression capabilities.

\begin{table}[!h]
	\centering
	\renewcommand\arraystretch{1.2}
	\begin{tabular}{c p{7cm}}
		\hline
		\hline
		\textbf{1.} & The $i$-th child node select two sets of Gaussian distributed random sequences $x_i$ and $p_i$, with length $n$, mean 0, and variance of $V_i$. Based on these, $n$ coherent states $\left|x_i+jp_i \right\rangle e^{j\omega_i t}$ are prepared by the $i$-th child node, and then register its own frequency band with the center node. Eventually, the $n$ coherent states generated by $n$ child nodes are transmitted to the center node through a quantum channel. \\
		\hline	
		\textbf{2.} & After receiving the multi-band coherent state $\sum_{i=1}^{n} \left|x_i+jp_i \right\rangle e^{j\omega_i t}$, the center node separates the coherent state $\left|x_i+jp_i \right\rangle e^{j\omega_i t}$ of the $i$-th child node using a filter.  Then, the center node eliminates the carrier through coherent demodulation to obtain $x_i$ and $p_i$ of the $i$-th child node. Finally, the coherent state modulated by each child node can be acquired through the phase shift recovery using the pilot signal. \\
		\hline
		\textbf{3.} & Because heterodyne detection is adopted, every child node retains all the data. \\
		\hline
		\textbf{4.} & Each child node acts as Alice and the center node acts as Bob to perform the same step 4 as shown in Tab. \ref{tab1}. \\
	    \hline
	    \textbf{5.} & Each child node acts as Alice and the center node acts as Bob to perform the same step 5 as shown in Tab. \ref{tab1}. \\
		\hline
		\hline
	\end{tabular}
	\caption{The steps of point-to-multipoint GMCS CV-QKD.}
	\label{tab2}
\end{table}

\subsubsection*{b. Distributed sensing integrated in quantum networks}

\begin{figure}[htb]
\centering
\subfigure[]{\label{fig2a}\includegraphics[width=1\linewidth]{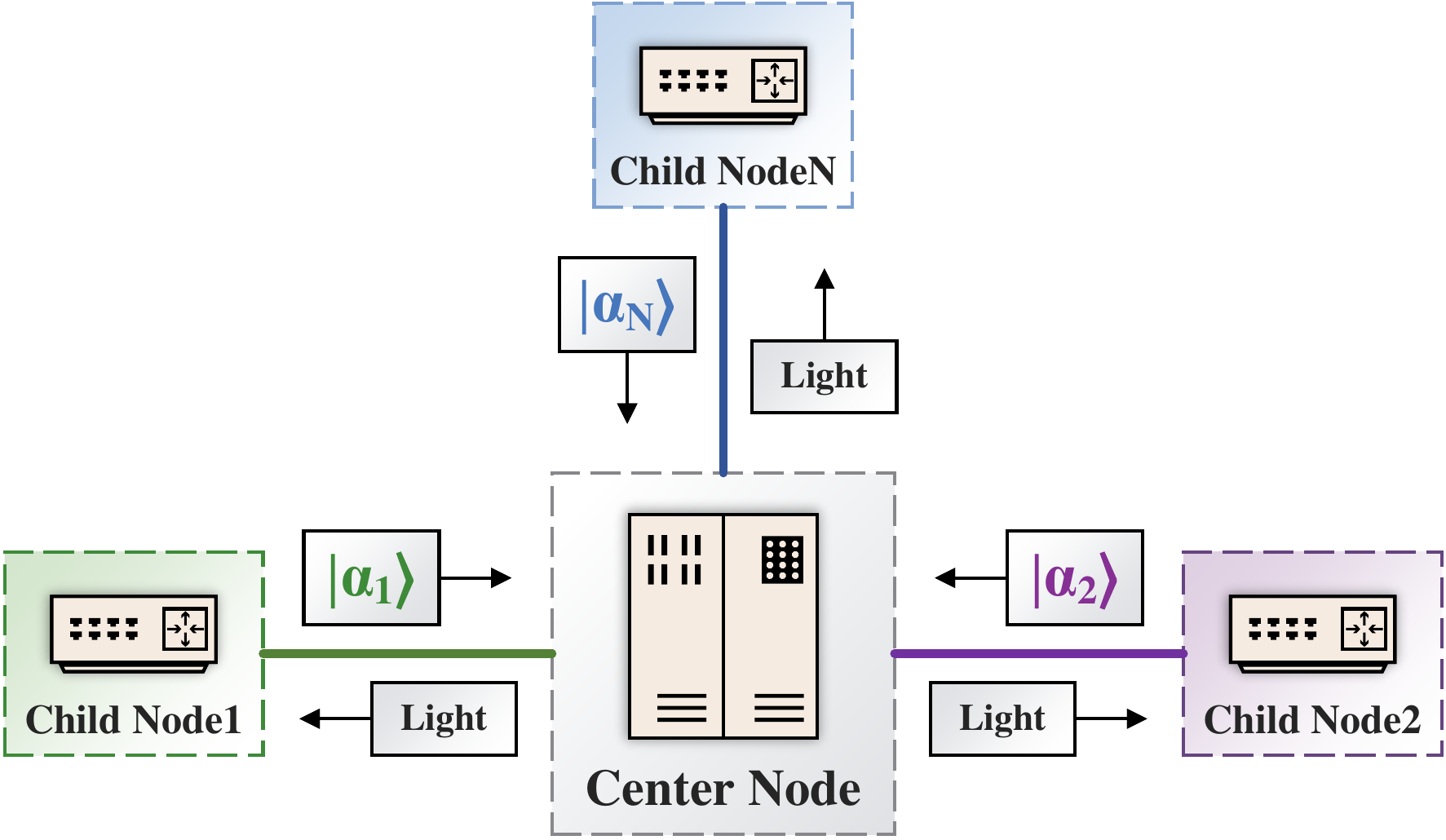}}
\subfigure[]{\label{fig2b}\includegraphics[width=1\linewidth]{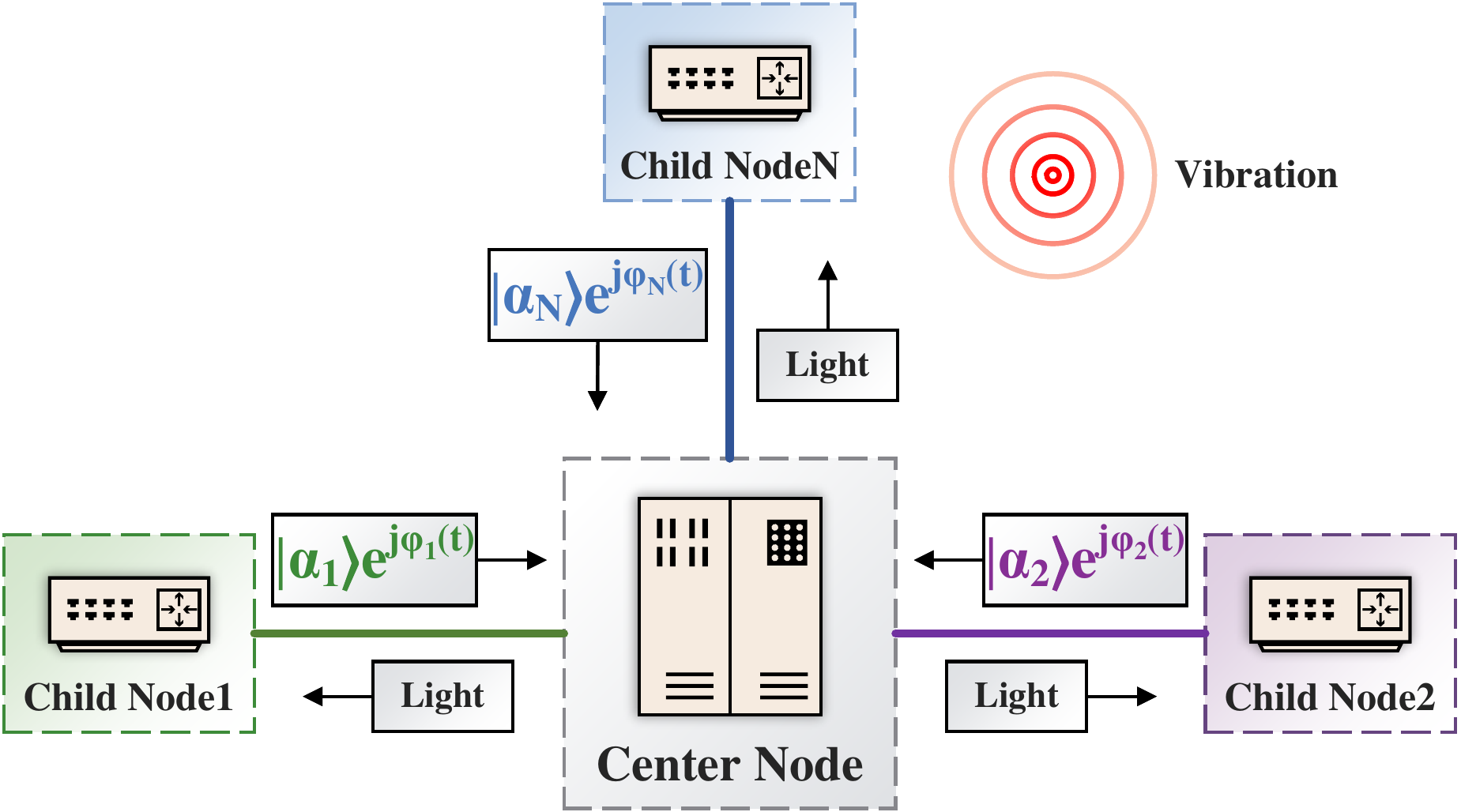}}
\caption{Schematic diagram of integrated sensing and QKD network (ISAQN) with and without vibration events. (a) When there are no vibration events, point-to-multipoint QKD operates normally. Each child node modulates its coherent state $\left|\alpha_{i}\right\rangle$ and returns it to the center node. (b) When vibration events occur, the coherent state of the $i$-th child node $\left|\alpha_{i}\right\rangle$ will rotate at an angle of $\varphi_i (t)$ and transform into $\left|\alpha_{i}\right\rangle e^{j\varphi_i \left(t\right)}$.}
\label{fig2}
\end{figure}

\begin{figure*}[htb]
\centering
\includegraphics[width=1\linewidth]{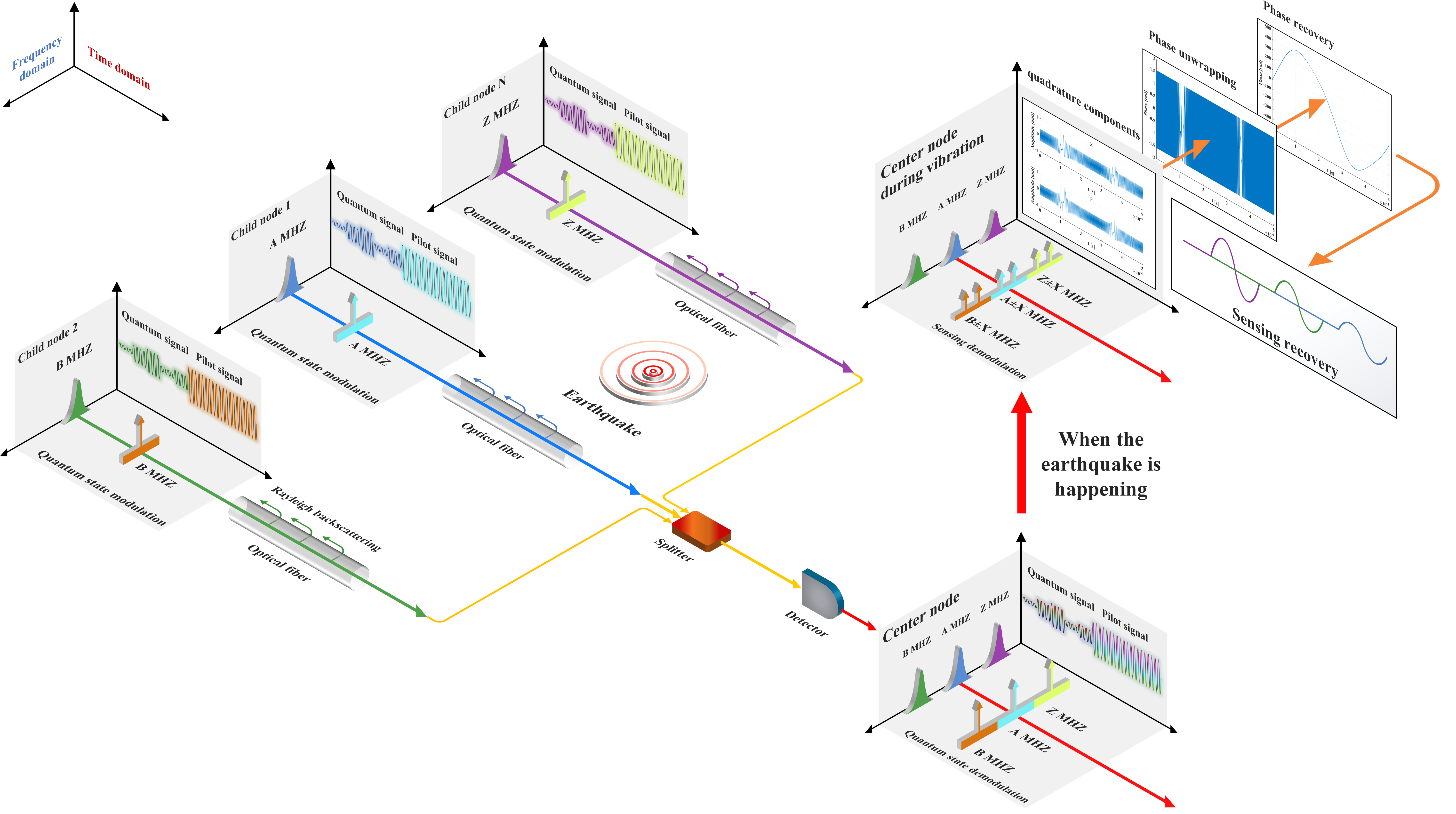}
\caption{Schematic diagram of ISAQN in the frequency domain and time domain. Different child node modulates quantum signals and pilot signals with different carrier frequencies. Then, they are converged through the optical fiber to the splitter and enter the detector for coherent detection. When the modulated coherent state and pilot signals do not experience fiber vibration events, the detected results remain a normal multi-band signal. When a vibration event occurs, the spectrum of the detected results experiences spectrum castdown due to the enhancement of Rayleigh backscattering, and the spectrum of the pilot experiences spectrum splitting in its centerline. The detection results are processed using the spectrum phase monitoring (SPM) protocol. By employing coherent demodulation, arctangent, and phase unwrapping, phase recovery can be accomplished, thereby obtaining the vibration waveform.}
\label{fig3}
\end{figure*}

DOFS technology achieves sensing by assessing the impact of environmental influences on various parameters of optical fibers. It can be categorized into DVS, DAS, and DTS based on its functionalities \cite{taylor1993apparatus,pan2011phase}. Classical DOFS usually uses back-scattering light to achieve sensing, which is also known as $\phi$-OTDR. It is mainly affected by the length of the optical fiber. The longer the optical fiber, the weaker the reflected back-scattering light, which affects sensing effectiveness. Therefore, this method requires strong optical power and usually uses an erbium-doped fiber amplifier (EDFA) to enhance the optical power \cite{juarez2005distributed,selker2006distributed,zhang2008distributed,koyamada2009fiber,lu2010distributed,lu2019distributed}, which is impossible in QKD.

Although the optical power of QKD is too weak to use the $\phi$-OTDR method, the QKD system can achieve sensing through the forward-transmitting light. Recently, the DOFS scheme using the forward-transmitting light has detected earthquakes in New Zealand and Japan using underwater fiber \cite{marra2018ultrastable}. In addition, DVS using the frequency locking link of QKD has also been proven to be feasible \cite{chen2022quantum}. The scheme requires sensitivity to phase variations and the ability to demodulate phase changes in the signal. It is mainly affected by the frequency-locking effect of the two lasers.

Through the above description of QKD, it is known that QKD signals are highly sensitive to the phase $\theta$. By utilizing the pilot signal within the QKD system, it becomes feasible to demodulate the phase $\theta$ and consequently restore the QKD signal. When this QKD system is affected by the environment and the phase $\theta$ changes, the demodulation and analysis of the sensing phase turn it into an ISACQ system. Furthermore, the $\theta_i$ of the $i$-th child node in different frequency bands can be distinguished in the round-trip multi-band QKD network. If each child node is considered as a sensing element, this network enables simultaneous sensing and QKD on the shared infrastructure, forming an ISAQN illustrated as Fig. \ref{fig2}. As shown in Fig. \ref{fig2a}, the ISAQN system functions normally when no vibration event occurs. The center node transmits light to each child node, and each child node transmits their respective coherent state $\left|\alpha_{i}\right\rangle$ back to the center node. When a vibration event is occurring, the coherent state $\left|\alpha_{i}\right\rangle$ modulated by the $i$-th child node will rotate at an angle of $\varphi_i\left(t\right)$, shown as Fig. \ref{fig2b}. At this moment, the coherent state modulated by the $i$-th child node becomes $\left|\alpha_{i}\right\rangle e^{j\varphi_i \left(t\right)}$. By extracting $\varphi_i(t)$ through pilot signals, the waveform of each child node's vibration can be reconstructed. The position and magnitude of the vibration event can be obtained by joint calculation of the vibration waveforms from all child nodes.

In the following, we will provide a detailed description of the specific principles for implementing the ISAQN. The phase delay $\varphi$ of the light transmitted through the optical fiber is
\begin{equation}
\begin{aligned}
\varphi=\beta L,
\end{aligned}
\end{equation}
where $\beta$ is the propagation constant and $L$ is the length of the optical fiber. When the optical fiber at the child node senses a vibration event, it withstands axial stress and radial stress. This will change its length $L$, refractive index $n$, and core diameter $D$, thus causing a change in the optical phase. Each child node experiences a different phase change. The expression for the phase change is
\begin{equation}
\begin{aligned}
\Delta \varphi &= \beta \Delta L+\Delta \beta L \\
&=\beta \Delta L+L \frac{d \beta}{d n} \Delta n+L \frac{d \beta}{d D} \Delta D,
\end{aligned}
\end{equation}
where $\Delta L$ is the change in optical fiber length, $\Delta n$ denotes the change in refractive index, and $\Delta D$ represents the change in core diameter. The first term describes the strain effect caused by the change in optical fiber length due to axial stress. The second term represents the photoelastic effect caused by the change of refractive index in the optical fiber. The third term is the Poisson effect caused by the change in optical fiber diameter due to radial stress. Since the Poisson effect is very small (usually $0.2\%$ of the photoelastic effect), it can be ignored. In addition, $\beta$ and $\Delta n$ can be expressed as
\begin{equation}
\begin{aligned}
\beta &=\frac{2 \pi n}{\lambda} \\
\Delta n &=-\frac{1}{2} n^3 \varepsilon_z\left[(1-\mu) p_{12}-\mu p_{11}\right],
\end{aligned}
\end{equation}
where $\lambda$ is the wavelength of light, $\varepsilon_z$ denotes the axial strain coefficient of the optical fiber, and has a relationship with the length change $\Delta L$ given by $\Delta L = \varepsilon_z L$. $\mu$ repesents the Poisson's ratio of the fiber, and $p_{11}$ and $p_{12}$ are the photoelastic constants of the fiber. Therefore, the phase change of the $i$-th child node is given by
\begin{equation}
\begin{aligned}
\Delta \varphi_i=\Delta L_i \frac{2 \pi}{\lambda}\left\{n-\frac{1}{2} n^3\left[(1-\mu) p_{12}-\mu p_{11}\right]\right\},
\end{aligned}
\label{formula11}
\end{equation}
where $\Delta L_i$ denotes the length change of the optical fiber in the $i$-th child node. It can be observed that the length change $\Delta L_i$ is directly proportional to the phase change $\Delta \varphi_i$. According to the above formula, the sensing effect can be amplified by changing the parameters of the optical fiber at the child nodes, thus making the child nodes serve as sensing elements. In practice, the long-distance fiber is placed in seismic-resistant pipelines, while the child nodes are placed in buildings. Therefore, the optical fiber in the child nodes is more sensitive to a vibration event compared to the long-distance fiber channels.

The schematic diagram of ISAQN in the time and frequency domains is shown in Fig. \ref{fig3}. Each child node modulates its quantum signal and pilot signal by TDM method according to Eq. \ref{formula1}. The optical power of the pilot signal is stronger than that of the quantum signal. The splitter collects the coherent states of all the child nodes and directs them to a coherent detector. The quantum signals from different users received by the detector do not interfere with each other in the frequency domain, and the same applies to the pilot signal. The constructed ISAQN is capable of sensing vibration events, such as an earthquake illustrated in Fig. \ref{fig3}. When the vibration waveform changes over time, the phase change and length change of the $i$-th child node $\Delta \varphi_i$ and $\Delta L_i$ becomes the time-varying $\varphi_i (t)$ and $L_i (t)$. The coherent state $\left|\alpha_{i}\right\rangle$ modulated by the $i$-th child node will rotate at an angle of $\varphi_i (t)$. At this point, the coherent state modulated by the $i$-th child node is derived from
\begin{equation}
\begin{aligned}
	\left|\alpha_{i}\right\rangle e^{j \varphi_{i} (t)} = \left|x_{i}+jp_{i} \right\rangle e^{j \left(\omega_i t + \varphi_{i} (t)\right)},i\in\{1,2,\cdots,N \}.
\end{aligned}
\end{equation}
As shown in Fig. \ref{fig3}, both the quantum signal and the pilot signal will result in the registered spectrum with spectrum castdown, indicating the perception of vibrations. This phenomenon is caused by the change in the refractive index of the optical fiber, which enhances the back-scattering light and reduces the power of forward-transmitting light. Due to the unique expression form of the pilot signal, it will also exhibit a phenomenon of spectrum splitting. By observing the location of the castdown or the splitting in the spectrum, we can determine which child nodes are experiencing vibration. The phase change of the $i$-th child node $\varphi_i (t)$ induced by vibration event can be demodulated through the coherent demodulation, which can be expressed as
\begin{equation}
\begin{aligned}
	&\left|Re\left(\alpha_{i} e^{j \varphi_i (t)}\right) \right\rangle \cos \left(\omega_i t \right) \\
	&=\left| \frac{1}{2} \left(\cos \left(2\omega_i t \right)+1\right)\left(x_{i} \cos \varphi_i (t) - p_{i} \sin \varphi_i (t)\right) \right.\\
	&\left. - \frac{1}{2}\sin \left(2\omega_i t \right)\left(x_{i} \sin \varphi_i (t) + p_{i} \cos \varphi_i (t) \right) \right\rangle, \\
	&\left|Im\left(\alpha_{i} e^{j \varphi_i (t)}\right) \right\rangle \cos\left(\omega_i t \right) \\
	&=\left| \frac{1}{2} \left(\cos \left(2\omega_i t \right)-1\right)\left(x_{i} \sin \varphi_i (t) + p_{i} \cos \varphi_i (t)\right)\right.\\
	&\left.+ \frac{1}{2}\sin \left(2\omega_i t \right)\left(x_{i} \cos \varphi_i (t) - p_{i} \sin \varphi_i (t) \right) \right\rangle.
\end{aligned}
\end{equation}
By applying a low-pass filter, the high-frequency components are effectively filtered out. As a result, we can obtain the quadrature components
\begin{equation}
\begin{aligned}
	\left|X_i\right\rangle=\left|x_{i} \cos \varphi_{i} (t) - p_{i} \sin \varphi_{i} (t) \right\rangle,\\
	\left|P_i\right\rangle=\left|x_{i} \sin \varphi_{i} (t) + p_{i} \cos \varphi_{i} (t) \right\rangle.
\end{aligned}
\end{equation}
Through substituting $x_i=1$ and $p_i=0$ of pilot signal into the above equation, the phase change of the $i$-th child node can be obtained by
\begin{equation}
\begin{aligned}
	\varphi_i \left(t\right) = \arctan \left( \left| X_i \right\rangle / \left| P_i \right\rangle \right).
\end{aligned}
\end{equation}
However, due to the limited value range of the arctangent function, which is constrained between $(-\pi, \pi)$, the phase change appears as a wrapped phase. In cases where the practical phase change exceeds this range, the demodulation results exhibit jumps at $\pm \pi$, with an amplitude of $2\pi$. This is known as the phase wrapping effect. Therefore, for phase variations caused by vibrations, we cannot directly observe regular waveforms such as sine waves in the demodulation results. Instead, we can only observe signals that vary in frequency over time, as shown in Fig. \ref{fig3}. By employing phase unwrapping, the demodulation results can be restored to the correct vibration waveform. The principle of phase unwrapping is to expand the value range of demodulated results. By detecting the differences between phases and adding or subtracting integer multiples of $2\pi$, the phase becomes continuously changing, which can be expressed as
\begin{equation}
\begin{aligned}
	\phi_i \left(t\right)=\varphi_i \left(t\right)+2 \pi m, \varphi_i \in \left(-\pi,\pi\right), m= 0, \pm 1, \pm 2 ,\cdots.
\end{aligned}
\end{equation}
As shown in Fig. \ref{fig3}, the result of phase unwrapping is a stable and continuous sinusoidal waveform, which matches the seismic vibration waveform that we set. Through experimental calibration or Eq. \ref{formula11}, it can be obtained that the length change of vibration is
\begin{equation}
\begin{aligned}
L_i \left(t\right) = \phi_i \left(t\right) \frac{\lambda}{2 \pi}{\left\{n-\frac{1}{2} n^3\left[(1-\mu) p_{12}-\mu p_{11}\right]\right\}}^{-1},
\end{aligned}
\end{equation}
The functionality of ISAQN sensing can be achieved by monitoring the spectrum and phase. Therefore, we refer to this sensing protocol as the SPM protocol. The SPM protocol can be summarized in Tab. \ref{tab3}.

\begin{table}[!h]
	\centering
	\renewcommand\arraystretch{1.2}
	\begin{tabular}{c p{7cm}}
		\hline
		\hline
		\textbf{1.} & When the point-to-multipoint GMCS CV-QKD in Tab. \ref{tab2} is operating normally, the center node monitors the frequency spectrum of quantum signals and pilot signals from all the child nodes. If the spectrum of the quantum signals and pilot signals from the $i$-th child node experience spectrum castdown, or if the spectrum of the pilot signal experiences spectrum splitting, it can be inferred that the $i$-th child node is perceiving vibrations. \\
		\hline	
		\textbf{2.} & If the vibration frequency and amplitude are low, the phase of the QKD signal can still be recovered by the pilot signal. In this case, the protocol proceeds directly to step 3. Otherwise, if the vibration frequency and amplitude are too high, the phase of the QKD signal can no longer be recovered from the pilot signal. At this point, the center node instructs the $i$-th child node to stop modulating the QKD signal and modulate only the pilot signal. \\
		\hline
		\textbf{3.} & The center node performs coherent demodulation on the pilot signal of the $i$-th child node to obtain quadrature components. Subsequently, the arctangent operation is applied to the quadrature components. Then, all phase jumps are eliminated in the results by employing phase unwrapping. Finally, the center node obtains the vibration waveform sensed by the $i$-th child node. \\
		\hline
		\hline
	\end{tabular}
	\caption{The steps of SPM protocol.}
	\label{tab3}
\end{table}

In the following, we will calculate the quantum limit that the measurement precision $\delta \varphi_i$ and $\delta L_i$ of sensing phase $\varphi_i$ and sensing length $L_i$ can achieve \cite{giovannetti2011advances}. Assuming the average photon numbers of the signal and the local oscillator (LO) are $\mathrm{N}_{\mathrm{s}}$ and $\mathrm{N}_{\mathrm{L}}$,  their annihilation operators are denoted as $\hat{\mathrm{a}}_{\mathrm{s}}$ and $\hat{\mathrm{a}}_{\mathrm{L}}$. After heterodyne detection, the annihilation operators of the $i$-th child node are represented as
\begin{equation}
\begin{aligned}
\hat{\mathrm{a}}_1=\frac{1}{2}\left(\hat{\mathrm{a}}_{\mathrm{s}} e^{j \varphi_i}+\hat{\mathrm{a}}_{\mathrm{L}} e^{j {\theta}_{\mathrm{sL}}}\right), \\
\hat{\mathrm{a}}_2=\frac{1}{2}\left(\hat{\mathrm{a}}_{\mathrm{s}} e^{j \varphi_i}-\hat{\mathrm{a}}_{\mathrm{L}} e^{j {\theta}_{\mathrm{sL}}}\right), \\
\end{aligned}
\end{equation}
where ${\theta}_{\mathrm{sL}}=\pi/2$ is the phase difference between LO and the signal, and $\hat{\mathrm{a}}$ is the annihilation operator. Then, the average photon numbers of $\hat{\mathrm{a}}_1$ and $\hat{\mathrm{a}}_2$ can be denoted as
\begin{equation}
\begin{aligned}
\hat{\mathrm{n}}_1 & ={\hat{\mathrm{a}}_1}^{\dagger} \hat{\mathrm{a}}_1 \\
& =\frac{1}{4}\left({\hat{\mathrm{a}}_{\mathrm{s}}}^{\dagger} \hat{\mathrm{a}}_{\mathrm{s}}+{\hat{\mathrm{a}}_{\mathrm{L}}}^{\dagger} \hat{\mathrm{a}}_{\mathrm{L}} \right. \\
& \left. +{\hat{\mathrm{a}}_{\mathrm{s}}}^{\dagger} \hat{\mathrm{a}}_{\mathrm{L}} e^{j\left(-\varphi_i+\theta_{\mathrm{sL}}\right)}+{\hat{\mathrm{a}}_{\mathrm{L}}}^{\dagger} \hat{\mathrm{a}}_{\mathrm{s}} e^{-j\left(-\varphi_i+\theta_{\mathrm{sL}}\right)} \right), \\
\hat{\mathrm{n}}_2 & ={\hat{\mathrm{a}}_2}^{\dagger} \hat{\mathrm{a}}_2 \\
& =\frac{1}{4}\left({\hat{\mathrm{a}}_{\mathrm{s}}}^{\dagger} \hat{\mathrm{a}}_{\mathrm{s}}+{\hat{\mathrm{a}}_{\mathrm{L}}}^{\dagger} \hat{\mathrm{a}}_{\mathrm{L}} \right. \\
& \left. -{\hat{\mathrm{a}}_{\mathrm{s}}}^{\dagger} \hat{\mathrm{a}}_{\mathrm{L}} e^{j\left(-\varphi_i+\theta_{\mathrm{sL}}\right)}-{\hat{\mathrm{a}}_{\mathrm{L}}}^{\dagger} \hat{\mathrm{a}}_{\mathrm{s}} e^{-j\left(-\varphi_i+\theta_{\mathrm{sL}}\right)} \right),
\end{aligned}
\end{equation}
where $\hat{\mathrm{a}}^{\dagger}$ is the creation operator. The average photon numbers after subtraction is
\begin{equation}
\begin{aligned}
\hat{\mathrm{n}}=\hat{\mathrm{n}}_1-\hat{\mathrm{n}}_2=\frac{1}{2} \left({\hat{\mathrm{a}}_{\mathrm{s}}}^{\dagger} \hat{\mathrm{a}}_{\mathrm{L}} e^{j\left(-\varphi_i+\theta_{\mathrm{sL}}\right)}+{\hat{\mathrm{a}}_{\mathrm{L}}}^{\dagger} \hat{\mathrm{a}}_{\mathrm{s}} e^{-j\left(-\varphi_i+\theta_{\mathrm{sL}}\right)} \right).
\end{aligned}
\end{equation}
Since LO is much stronger than the signal, we can obtain $\alpha_\mathrm{L}\gg\alpha_\mathrm{s}$. In the case of calculating precision, sensing phase $\varphi_i$ can be viewed as infinitesimal. For a very small phase $\varphi_i$, the annihilation operator of LO can be treated as a classical quantity, represented as ${\hat{\mathrm{a}}_{\mathrm{L}}}^{\dagger}=\hat{\mathrm{a}}_{\mathrm{L}}=\alpha_\mathrm{L}$. Therefore, we can get
\begin{equation}
\begin{aligned}
\hat{\mathrm{n}}=\frac{\alpha_{\mathrm{L}}}{2} \left({\hat{\mathrm{a}}_{\mathrm{s}}}^{\dagger} e^{j\left(-\varphi_i+\theta_{\mathrm{sL}}\right)}+\hat{\mathrm{a}}_{\mathrm{s}} e^{-j\left(-\varphi_i+\theta_{\mathrm{sL}}\right)}\right).
\end{aligned}
\end{equation}
By substituting ${\theta}_{\mathrm{sL}}=\pi/2$, the above formula can be simplified to
\begin{equation}
\begin{aligned}
\hat{\mathrm{n}}=\frac{j \alpha_{\mathrm{L}}}{2} \left({\hat{\mathrm{a}}_{\mathrm{s}}}^{\dagger} e^{-j\varphi_i}-\hat{\mathrm{a}}_{\mathrm{s}} e^{j \varphi_i}\right).
\end{aligned}
\end{equation}
Utilizing linearization formulas $\hat{\mathrm{a}}_{\mathrm{s}}=\alpha_{\mathrm{s}}+\delta \hat{\mathrm{a}}_{\mathrm{s}}$ and ${\hat{\mathrm{a}}_{\mathrm{s}}}^{\dagger}={\alpha_{\mathrm{s}}}^*+\delta {\hat{\mathrm{a}}_{\mathrm{s}}}^{\dagger}$, we can obtain the following equation
\begin{equation}
\begin{aligned}
\hat{\mathrm{n}} = \alpha_{\mathrm{L}}\left(\alpha_{\mathrm{s}} \sin (\varphi_i)-j\left(\delta \hat{\mathrm{a}}_{\mathrm{s}}-\delta \hat{\mathrm{a}}_{\mathrm{s}}^{\dagger}\right)\right).
\end{aligned}
\end{equation}
Since $\varphi_i$ is infinitesimal, it follows that $\sin (\varphi_i)=\varphi_i$. The real numbers $\alpha_\mathrm{L}$ and $\alpha_\mathrm{s}$ can be replaced by average photon number $\alpha_\mathrm{L}=\sqrt{N_\mathrm{L}}$ and $\alpha_\mathrm{s}=\sqrt{N_\mathrm{s}}$ to derive the following formula
\begin{equation}
\begin{aligned}
\hat{\mathrm{n}} = \sqrt{N_\mathrm{L}}\left(\sqrt{N_\mathrm{s}} \varphi_i-j\left(\delta \hat{\mathrm{a}}_{\mathrm{s}}-\delta \hat{\mathrm{a}}_{\mathrm{s}}^{\dagger}\right)\right).
\end{aligned}
\end{equation}
Thus, the quantum limit of sensing phase precision $\delta \varphi_i$ can be expressed as
\begin{equation}
\begin{aligned}
\delta \varphi_i=\frac{\delta \hat{\mathrm{Y}}_{\mathrm{s}}}{\sqrt{N_\mathrm{s}}},
\end{aligned}
\label{formula26}
\end{equation}
where $\delta \hat{\mathrm{Y}}_{\mathrm{s}}=j\left(\delta \hat{\mathrm{a}}_{\mathrm{s}}-\delta \hat{\mathrm{a}}_{\mathrm{s}}^{\dagger}\right)$ is the vacuum fluctuation. From this formula, it can be inferred that the sensing phase precision is related to the vacuum fluctuation $\delta \hat{\mathrm{Y}}_{\mathrm{s}}$ and the average photon number $N_\mathrm{s}$, which reaches the standard quantum limit. By substituting in Eq. \ref{formula26}, Eq. \ref{formula11} can be derived to
\begin{equation}
\begin{aligned}
\delta L_i=\frac{2 \pi \delta \hat{\mathrm{Y}}_{\mathrm{s}}}{\lambda \sqrt{N_\mathrm{s}}}\left\{n-\frac{1}{2} n^3\left[(1-\mu) p_{12}-\mu p_{11}\right]\right\},
\end{aligned}
\end{equation}
where $\delta L_i$ is the measurement precision of sensing length $L_i$. In conclusion, the detection precision of ISAQN can reach the standard quantum limit, which is consistent with the shot noise limit in the experiment. Although the precision of this sensing method cannot reach the Heisenberg quantum limit, it can achieve a standard quantum limit that is difficult to attain by other classical sensing methods.

\begin{figure}[htb]
\centering
\includegraphics[width=1\linewidth]{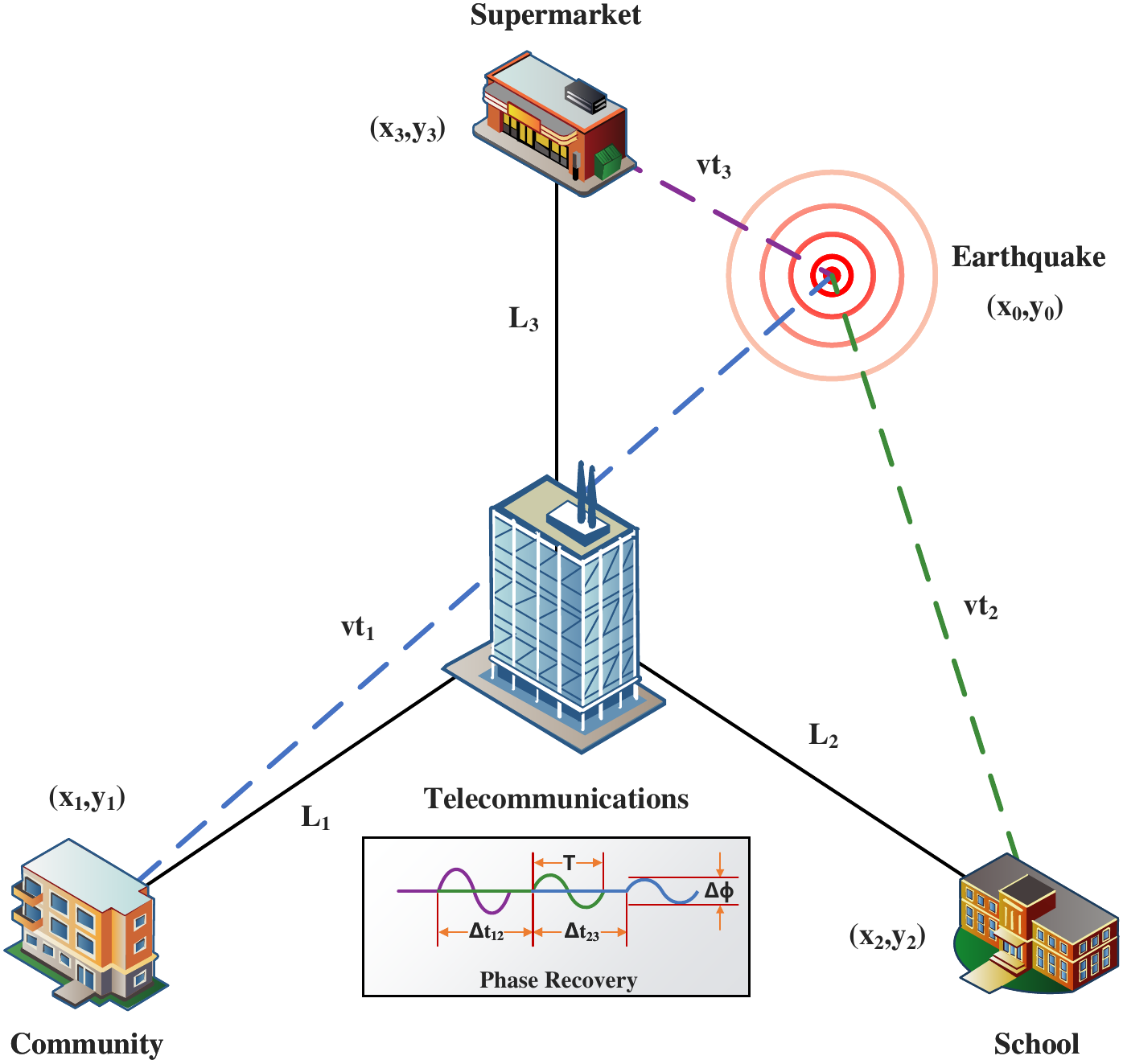}
\caption{Schematic diagram of calculating the earthquake's position and magnitude. When an earthquake occurs at $(x_0, y_0)$, it's detected by three child nodes at $(x_1, y_1)$, $(x_2, y_2)$, and $(x_3, y_3)$. Phase recovery depicts the vibration waveforms sensed by three child nodes. Due to their different positions, the arrival time $t_1$, $t_2$, $t_3$ and $\Delta \phi$ vary, but the period $T$ and waveforms remain the same. $\Delta t_{12}$ and $\Delta t_{23}$ are the time differences between the arrival of the waveforms. Assuming seismic wave propagation speed is $v$, the distances for the wave to reach nodes are $vt_1$, $vt_2$, and $vt_3$. Moreover, $L_1$, $L_2$, and $L_3$ represent distances from these nodes to the center node.}
\label{fig4}
\end{figure}

To illustrate how ISAQN facilitates distributed sensing of vibration events, we use an earthquake as an example to describe it. Assuming the seismic wave is denoted as $E(t)$, the phase detected by the $i$-th child node can be represented as
\begin{equation}
\begin{aligned}
	E_i(t) = \gamma_i E(t+t_i),
\end{aligned}
\end{equation}
where $\gamma_i \propto t_i$ is the attenuation coefficient when the seismic wave arrives at the $i$-th child node, and $t_i$ represents the time spent. Once the vibration waveforms of each child node are demodulated in the center node, the position and magnitude of the vibration can be calculated, as shown in Fig. \ref{fig4}. It is achieved by analyzing the vibration waveform with different arrival time $t_i$ or different attenuation coefficient $\gamma_i$. The time difference of the vibration waveforms arriving at different child nodes, obtained through cross-correlation, allows for the derivation of the position. The magnitude of the vibration experienced by each child node can be determined by the amount of phase change. This enables the reconstruction of the original vibration event, including its position and magnitude. At least three child nodes are required to complete this process. In the following, we will present a comprehensive explanation of the calculation process.

For example, assuming an earthquake occurs at position $(x_0, y_0)$, the child node located at position $(x_3, y_3)$ detects the vibration first, represented by the purple waveform in the phase recovery. Next, the child node located at position $(x_2, y_2)$ senses the vibration, represented by the green waveform. Finally, the child node located at position $(x_1, y_1)$ senses the vibration, represented by the blue waveform. The vibration waveforms of these three child nodes are plotted in Fig. \ref{fig4}, where $\Delta t_{12}$ represents the time difference between the first and second arriving vibration waveforms, and $\Delta t_{23}$ represents the time difference between the second and third arriving vibration waveforms. Assuming the propagation speed of the seismic wave is $v$, the time required for the wave to reach child node $(x_1, y_1)$, $(x_2, y_2)$, and $(x_3, y_3)$ are $t_1$, $t_2$, and $t_3$. In addition, $L_1$, $L_2$, and $L_3$ are used to represent the distances between the center node and child nodes $(x_1, y_1)$, $(x_2, y_2)$, and $(x_3, y_3)$, respectively. Due to the attenuation of seismic waves with increasing distance during propagation, the phase change $\Delta \phi$ and arrival time $t_1$, $t_2$, $t_3$ differ. However, the period $T$ and waveform remain the same. Therefore, the following equations can be obtained
\begin{equation}
\begin{aligned}
\left(x_{1}-x_{0}\right)^2 &+\left(y_{1}-y_{0}\right)^2=(v t_{1})^2, \\
\left(x_{2}-x_{0}\right)^2 &+\left(y_{2}-y_{0}\right)^2=(v t_{2})^2, \\
\left(x_{3}-x_{0}\right)^2 &+\left(y_{3}-y_{0}\right)^2=(v t_{3})^2, \\
t_{2}-t_{3}&=\Delta t_{12} + \left(L_{2}-L_{3}\right) /c, \\
t_{1}-t_{2}&=\Delta t_{23} + \left(L_{1}-L_{2}\right) /c,
\end{aligned}
\end{equation}
where $c$ is the speed of light in the optical fiber. The intersection of the three circles corresponds to a unique point, allowing for the determination of the vibration position. Since the phase change $\Delta \phi$ is proportional to the magnitude of the vibration, the original magnitude of the vibration can be determined by measuring the phase change at each child node.

In conclusion, the ISAQN achieves the coexistence of QKD and sensing. Without the requirement for additional devices, the round-trip multi-band QKD network can be transformed into an ISAQN through the SPM protocol. It can also sense the location and magnitude of a vibration event, such as an earthquake.

\begin{figure*}[htb]
\centering
\includegraphics[width=1\linewidth]{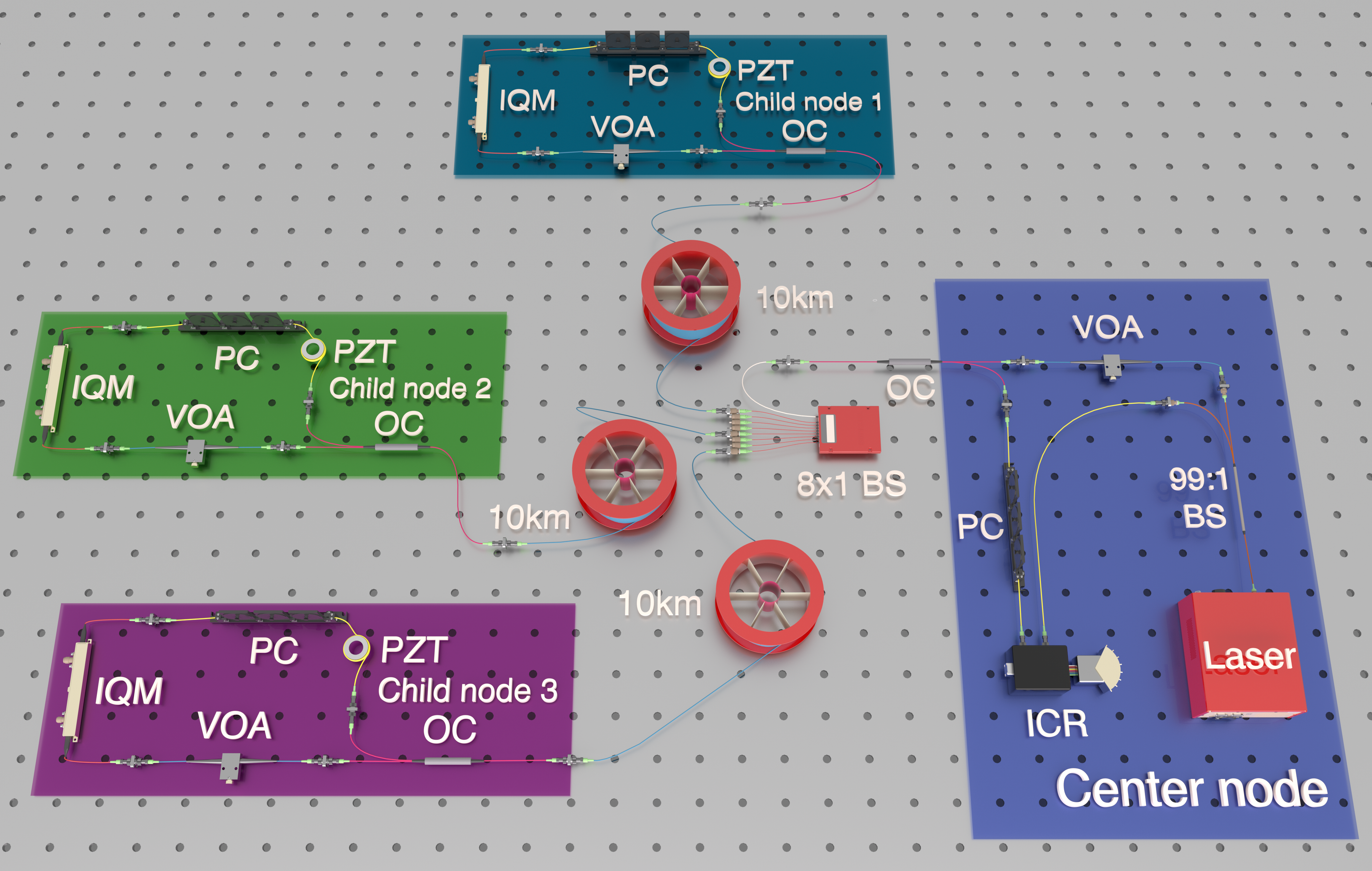}
\caption{Optical structure of ISAQN. The light from the center node is split by a 99:1 beam splitter (BS) into a high-power local oscillator (LO) and low-power light which passes through a variable optical attenuator (VOA) and an optical circulator (OC). VOA reduces light to the quantum level and the 8$\times$1 BS distributes it to eight child nodes. The network capacity of the ISAQN constructed in the experiment is 8. Due to experimental equipment limitations, only three nodes are used, each connected to a 10 $\rm{km}$ optical fiber. This light passes through another VOA for power balance, is modulated via an arbitrary waveform generator (AWG), and then reaches child nodes. Child nodes 1, 2, and 3 modulate 100, 200, and 300 $\rm{MHz}$ carrier frequencies respectively, with a baseband frequency of 50 $\rm{MHz}$. Then, the signal enters an optical fiber wound around the piezoelectric transducer (PZT), which is used to simulate vibration events. After passing through the 10 $\rm{km}$ optical fiber, the signals are merged back via the 8$\times$1 BS, and return to the center node where polarization is adjusted again before reaching the integrated coherent receiver (ICR).}
\label{fig5}
\end{figure*}

\subsection*{Experiment verification of ISAQN}

Based on the theoretical derivation of ISAQN, we conducted a proof-of-principle experiment. This experiment simultaneously implemented point-to-multipoint quantum networks and distributed sensing networks in ISAQN. Firstly, we will illustrate the experiment set-up in part a. Secondly, we will present a performance analysis for quantum communication and sensing in part b.

\subsubsection*{a. Experiment set-up}

\begin{figure*}[htb]
\centering
\subfigure[]{\label{fig6a}\includegraphics[width=0.3\linewidth]{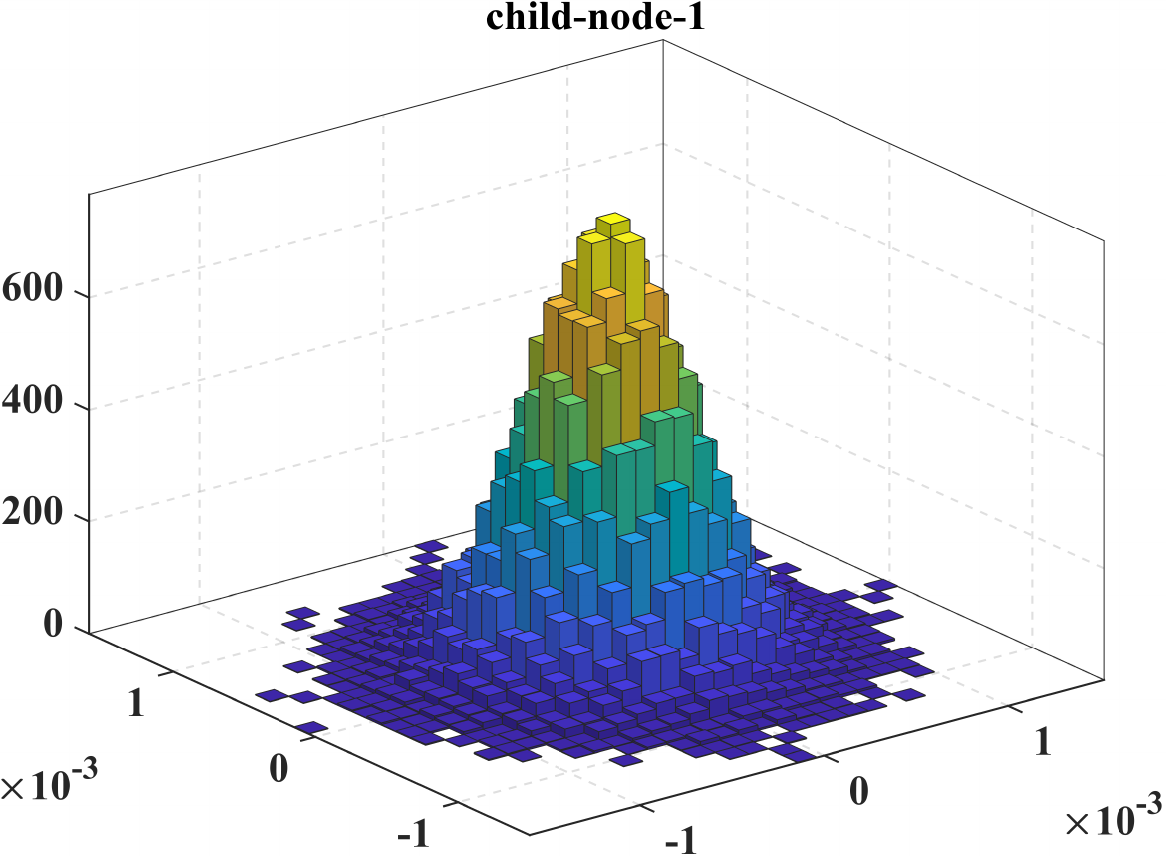}}
\hspace{3ex}
\subfigure[]{\label{fig6b}\includegraphics[width=0.3\linewidth]{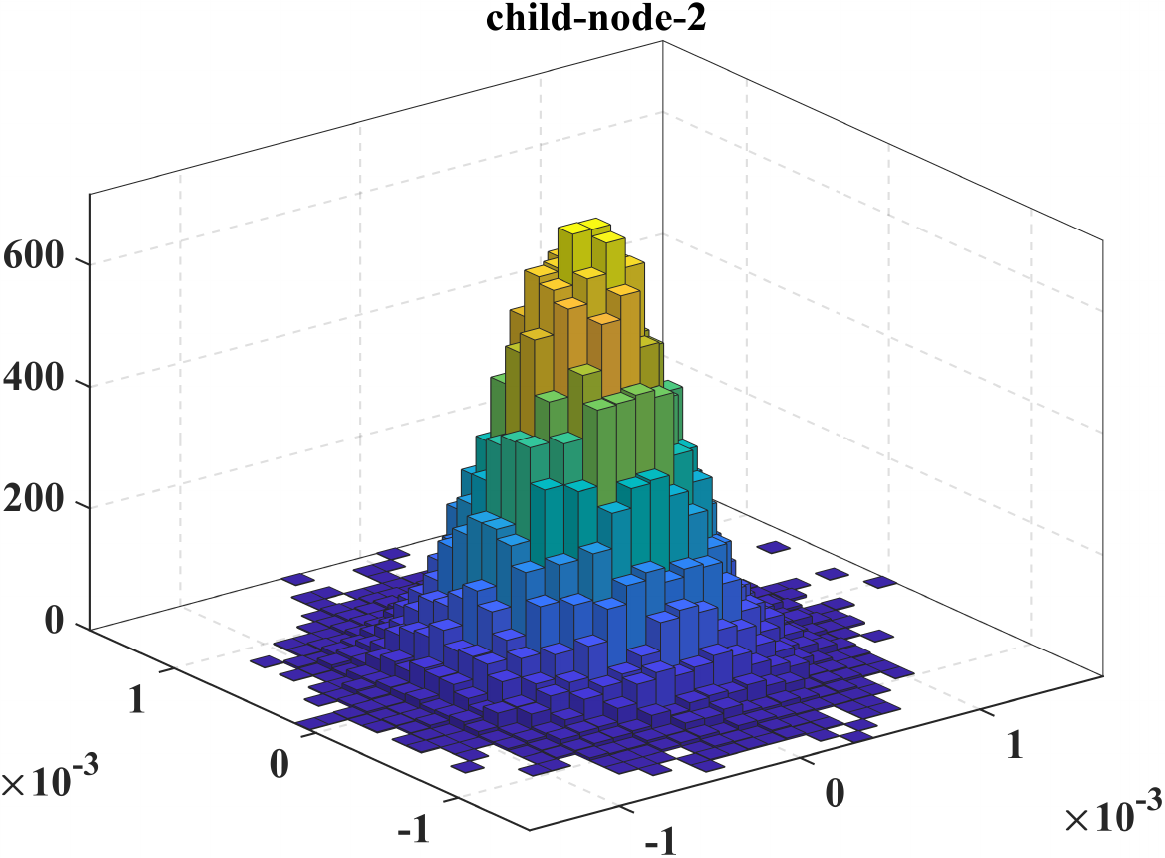}}
\hspace{3ex}
\subfigure[]{\label{fig6c}\includegraphics[width=0.3\linewidth]{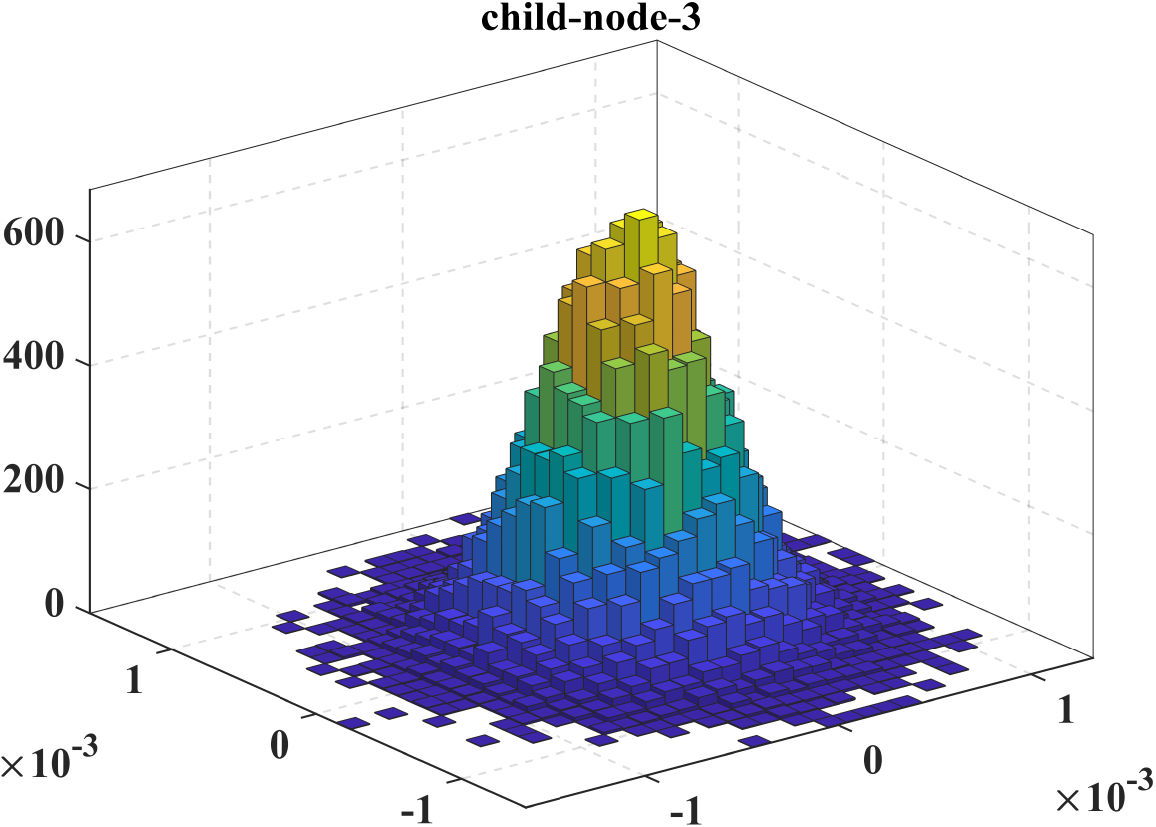}}
\caption{Distribution of experimental data in the phase space for three child nodes. The experimental data of each child node follows a Gaussian distribution.}
\label{fig6}
\end{figure*}

\begin{figure}[htb]
\centering
\includegraphics[width=1\linewidth]{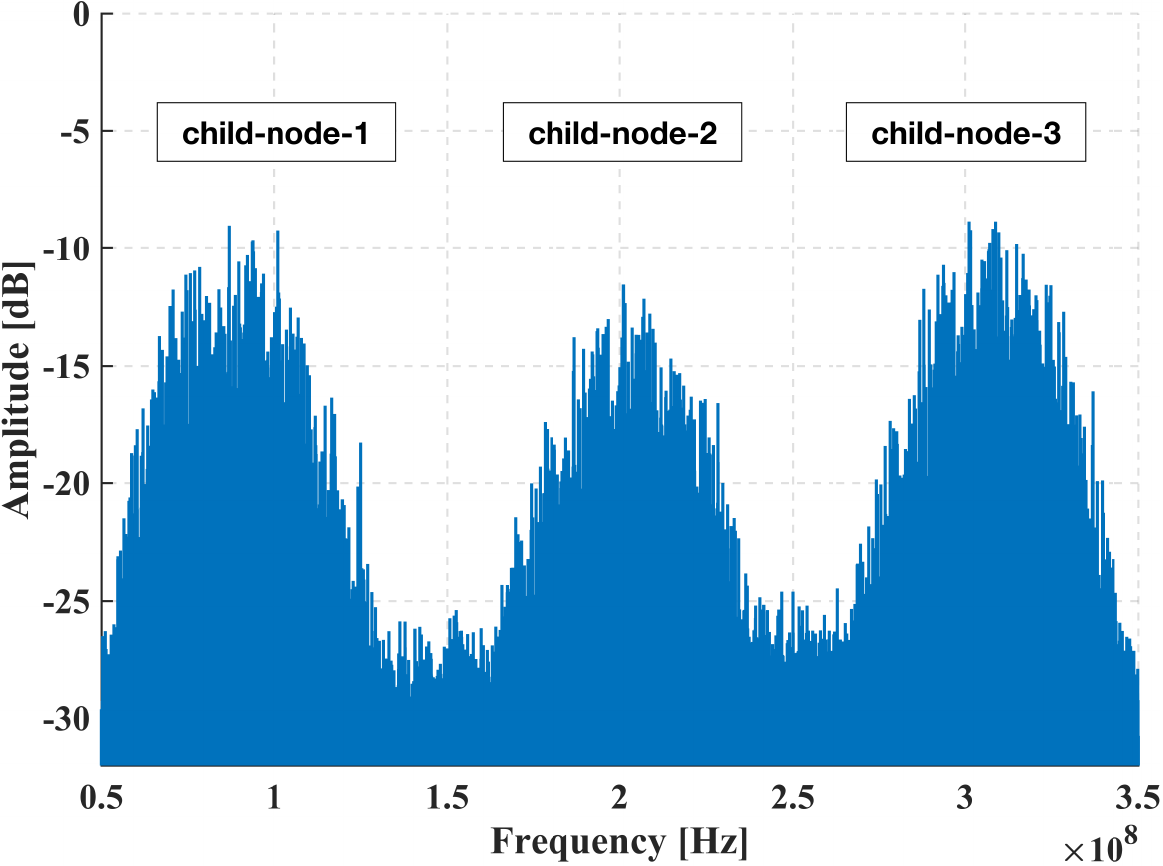}
\caption{Frequency spectrum of the signal received by the center node. There are three frequency bands of three child nodes, which are 100 $\rm{MHz}$, 200 $\rm{MHz}$, and 300 $\rm{MHz}$. When light attenuates to the quantum level, its band becomes difficult to observe. To enhance the visual clarity and distinction of the spectrum, we increased the optical power during the drawing process.}
\label{fig7}
\end{figure}

\begin{figure}[htb]
\centering
\includegraphics[width=1\linewidth]{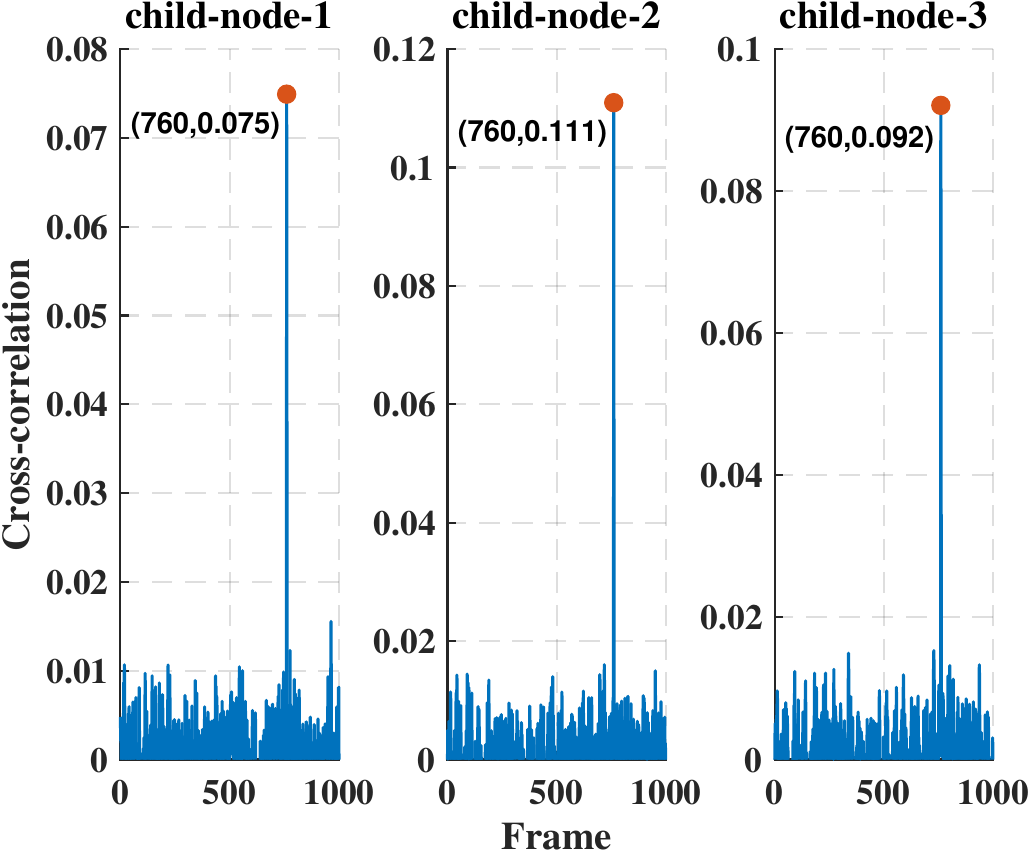}
\caption{Cross-correlation of three child nodes and center node. Since AWG and oscilloscope are clock-synchronized, the frame synchronization offset of each node is the same.}
\label{fig8}
\end{figure}

\begin{figure}[htb]
\centering
\includegraphics[width=1\linewidth]{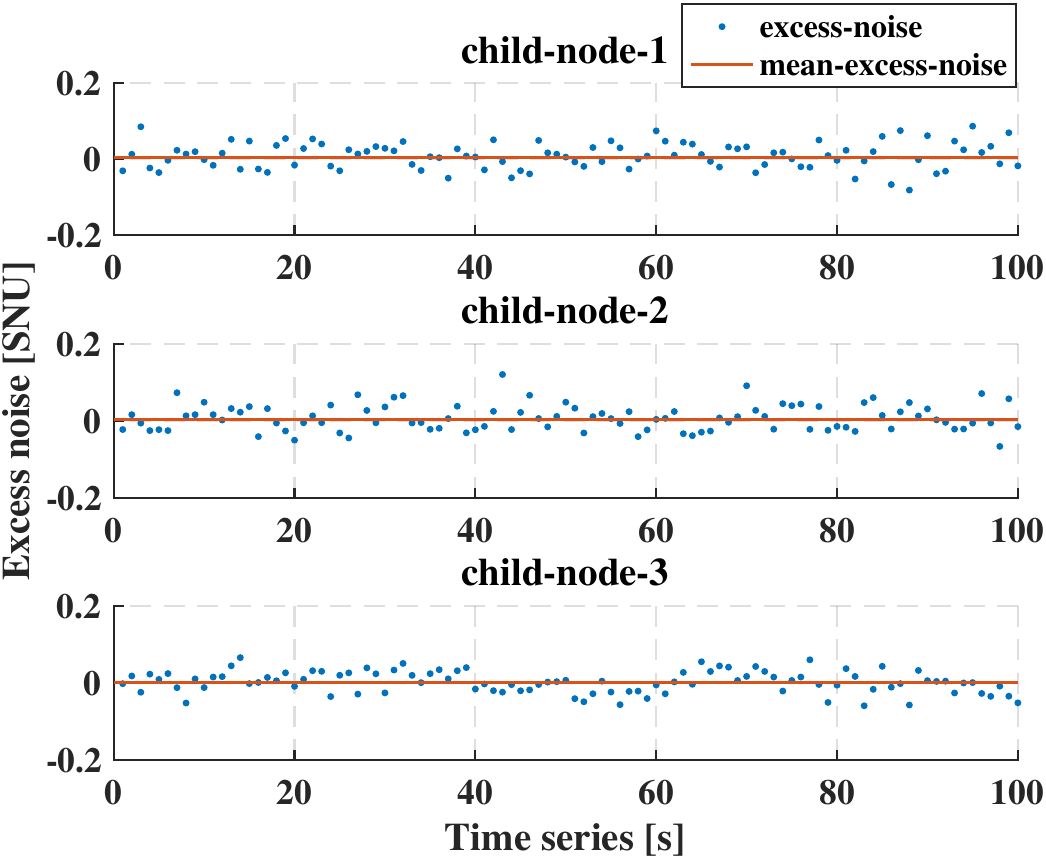}
\caption{Excess noise of three child nodes. $10^5$ points are collected per second to calculate the excess noise. The blue scattered points represent the excess noise at different times. The red straight line denotes the mean excess noise over 100 seconds. The excess noise means of three child nodes are 4.7 $\mathrm{mSNU}$, 2.4 $\mathrm{mSNU}$, and 3.6 $\mathrm{mSNU}$ respectively, where $\rm{SNU}$ is the shot noise unit.}
\label{fig9}
\end{figure}

\begin{figure}[htb]
\centering
\includegraphics[width=1\linewidth]{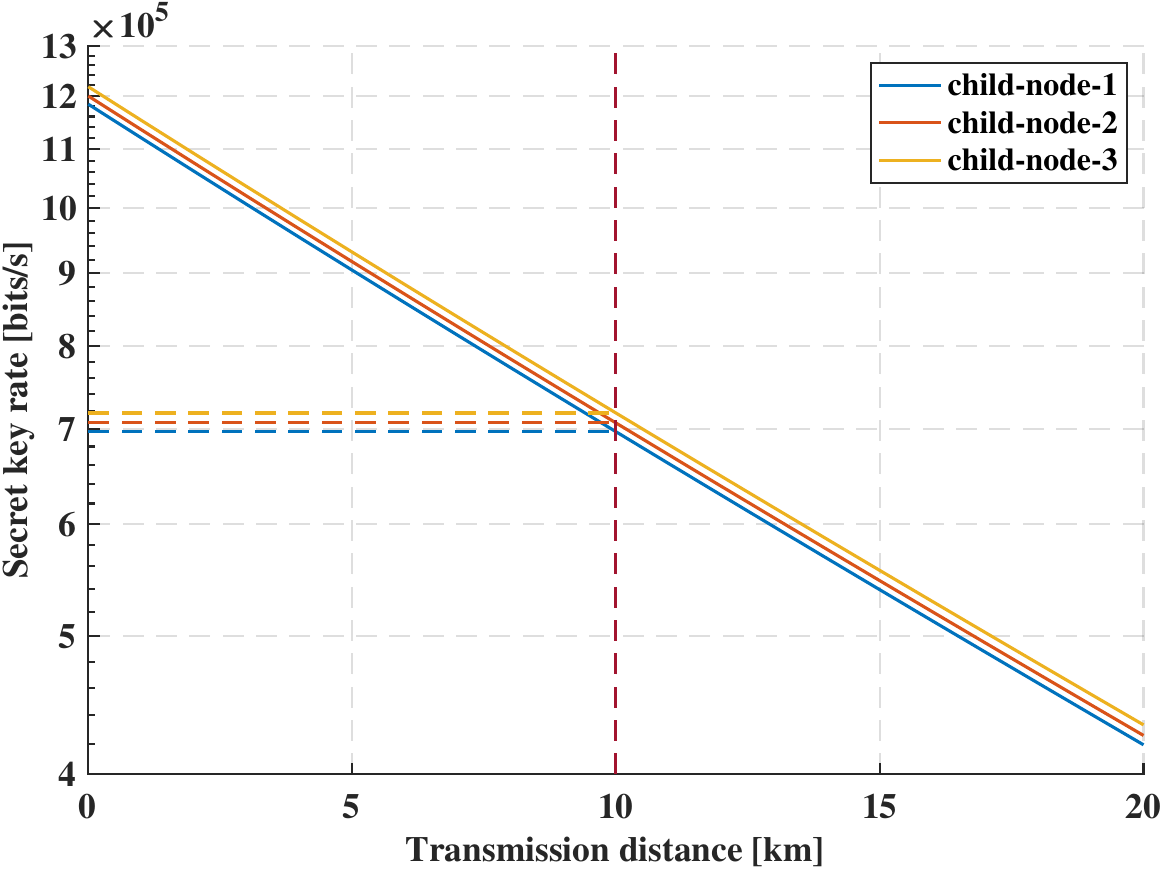}
\caption{Secret key rate (SKR) of three child nodes. As the transmission distance increases, the SKR of the three child nodes decreases. At the transmission distance $L=10$ $\rm{km}$, the SKR of three child nodes are respectively 0.70 $\rm{Mbits/s}$, 0.71 $\rm{Mbits/s}$, and 0.72 $\rm{Mbits/s}$.}
\label{fig10}
\end{figure}

\begin{figure}[!h]
\centering
\includegraphics[width=1\linewidth]{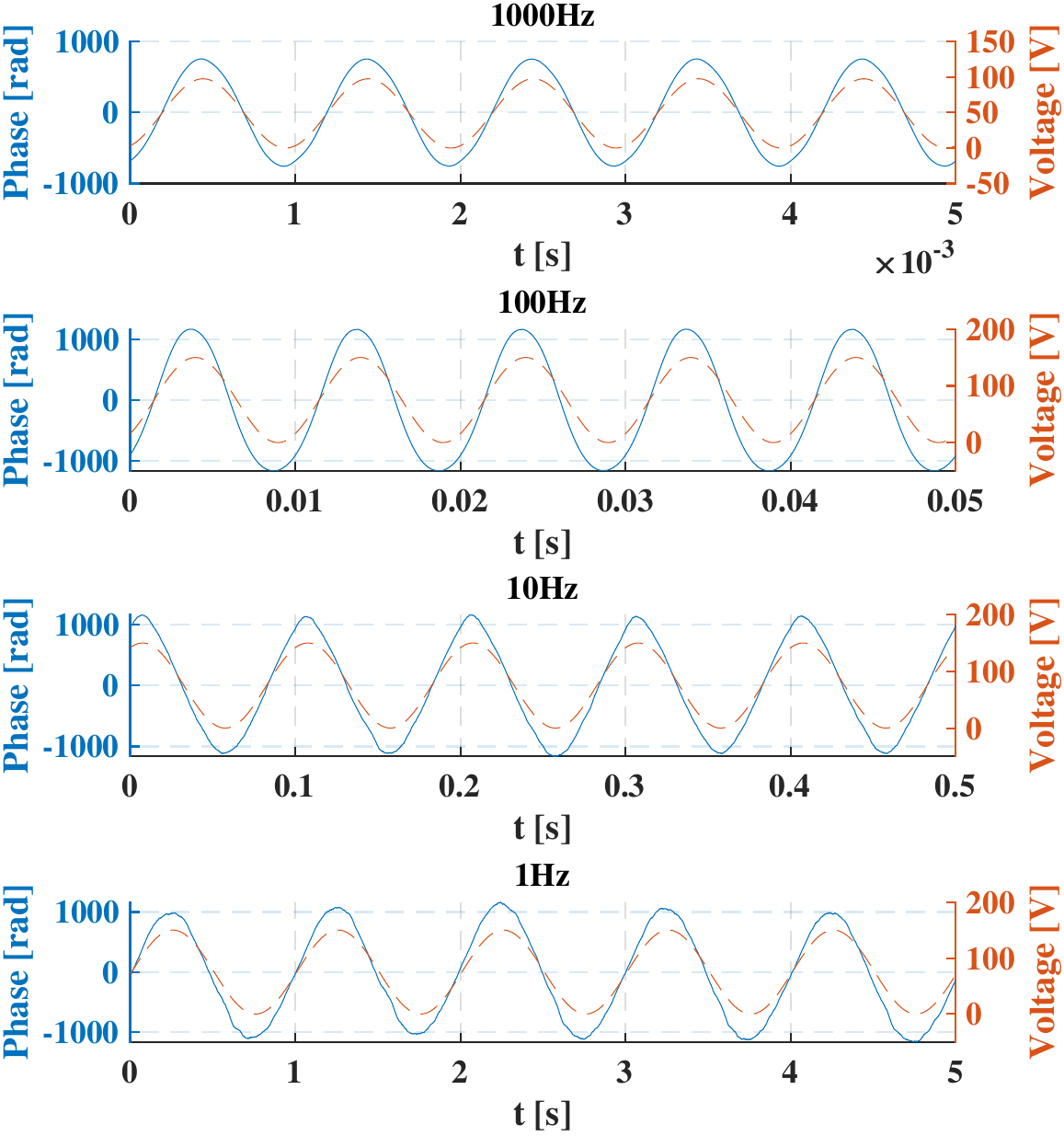}
\caption{Phase waveform recovered by ISAQN and voltage waveform loaded on PZT. The blue solid line represents the phase waveform, while the red dashed line represents the voltage waveform at the same time.}
\label{fig11}
\end{figure}

The optical structure of ISAQN used in this experiment is illustrated in Fig. \ref{fig5}. First, the light transmitted from the center node is divided into two beams of light by a 99:1 beam splitter (BS). The high-power light is used as the LO for coherent detection, while the low-power light passes through a variable optical attenuator (VOA) before reaching the optical circulator (OC). The VOA of the center node reduces the light to the quantum level. After the light is transmitted from port 1 to port 2 of the OC, it is evenly distributed to eight child nodes through an 8$\times$1 BS. The network capacity of the ISAQN constructed in the experiment is 8. However, due to experimental equipment limitations, only three child nodes were used in the experiment. Each child node is connected to a 10 $\rm{km}$ optical fiber. After reaching the child nodes, the light is transmitted from port 2 to port 3 of OC and then passes through another VOA. The VOA of the child node is used to balance the optical power. The optical signal is then modulated using an arbitrary waveform generator (AWG) through IQM. Child node 1, child node 2, and child node 3 modulate carrier frequencies of 100 $\rm{MHz}$, 200 $\rm{MHz}$ and 300 $\rm{MHz}$, respectively, while the baseband frequency for each child node is 50 $\rm{MHz}$. After modulation, the signal from each child node is adjusted for polarization using a polarization controller (PC). Then, the signal enters a 2.5 $\rm{m}$ optical fiber wound around the piezoelectric transducer (PZT), with dimensions of 53.10 $\rm{mm}$ $\times$ 55.00 $\rm{mm}$ $\times$ 3.95 $\rm{mm}$. In the experiment, different voltages are applied to the PZT to simulate different vibration events. Subsequently, the signal returns to the 10 $\rm{km}$ optical fiber from port 1 to port 2 of OC. After the signals of all child nodes pass through the 8$\times$1 BS once again, they return to the center node via ports 2 and 3 of OC. At the center node, the signals are integrally adjusted for polarization using a PC and then reach the integrated coherent receiver (ICR) together with LO. The detected signal is sampled by an oscilloscope.

\subsubsection*{b. Experiment result}

For QKD, the distributions of experimental data in the phase space for three child nodes are shown in Fig. \ref{fig6}. The modulation variance of each child node is set to $V_\mathrm{A}=V_1=V_2=V_3=12$ $\mathrm{SNU}$, where $\rm{SNU}$ is the shot noise unit. The spectrum of the signal received by the center node through ICR is presented in Fig. \ref{fig7}. The mixed spectrum received by the center node has frequencies of 100 $\rm{MHz}$, 200 $\rm{MHz}$, and 300 $\rm{MHz}$, and there is no occurrence of spectrum aliasing phenomenon. In order to make the spectrum more visual and distinct, we increased the optical power when drawing Fig. \ref{fig7}. The actual spectrum of the quantum signal is much smaller than what is shown in Fig. \ref{fig7}. Additionally, Fig. \ref{fig8} exhibits the results of the cross-correlation between the signals modulated by each child node and received by the center node. The cross-correlation results demonstrate the success of frame synchronization, as indicated by the prominent vertical lines that remain visible. The excess noise scatter plots and excess noise mean of 100 data frames with $10^5$ points from three child nodes are depicted in Fig. \ref{fig9}. The excess noise means of three child nodes are 4.7 $\mathrm{mSNU}$, 2.4 $\mathrm{mSNU}$, and 3.6 $\mathrm{mSNU}$ respectively. It can be inferred that ISAQN has excellent noise suppression capability.

In this experiment, we also evaluate the reachable SKR for GMCS CV-QKD. The formula of the SKR for unit system repetition frequency is in Appendix. For the practical CV-QKD system, SKR $K$ can be calculated as
\begin{equation}
\begin{aligned}
K = RK_r,
\end{aligned}
\end{equation}
where $R$ is the repetition frequency of the CV-QKD system. The other parameters introduced in the calculation are quantum efficiency $\eta = 0.42$, electrical noise $v_{el}=0.18$, reconciliation efficiency $\beta=0.98$, modulation variance $V_{\rm{A}}=12$ $\rm{SNU}$ and repetition frequency $R=50$ $\rm{MHz}$. The $N\times1$ BS on the return path would introduce a $1/N$ loss on each arm, thereby reducing the SKR of all child nodes. The transmittance will be changed to $T=10^{-\alpha L / 10} / N$, where $\alpha=0.2$ $\rm{dB/km}$ denotes the attenuation coefficient of optical fiber, and $N$ is the network capacity of ISAQN and branch number of the $N\times1$ BS. In the experiment, we substituted $N=8$ into the formula and obtained the secure key rates for three child nodes as shown in Fig. \ref{fig10}. At the transmission distance $L=10$ $\rm{km}$, the SKR of three child nodes are respectively 0.70 $\rm{Mbits/s}$, 0.71 $\rm{Mbits/s}$, and 0.72 $\rm{Mbits/s}$. In conclusion, ISAQN has achieved outstanding experimental results in multi-point QKD.

\begin{figure*}[htb]
\centering
\subfigure[]{\label{fig12a}\includegraphics[width=0.3\linewidth]{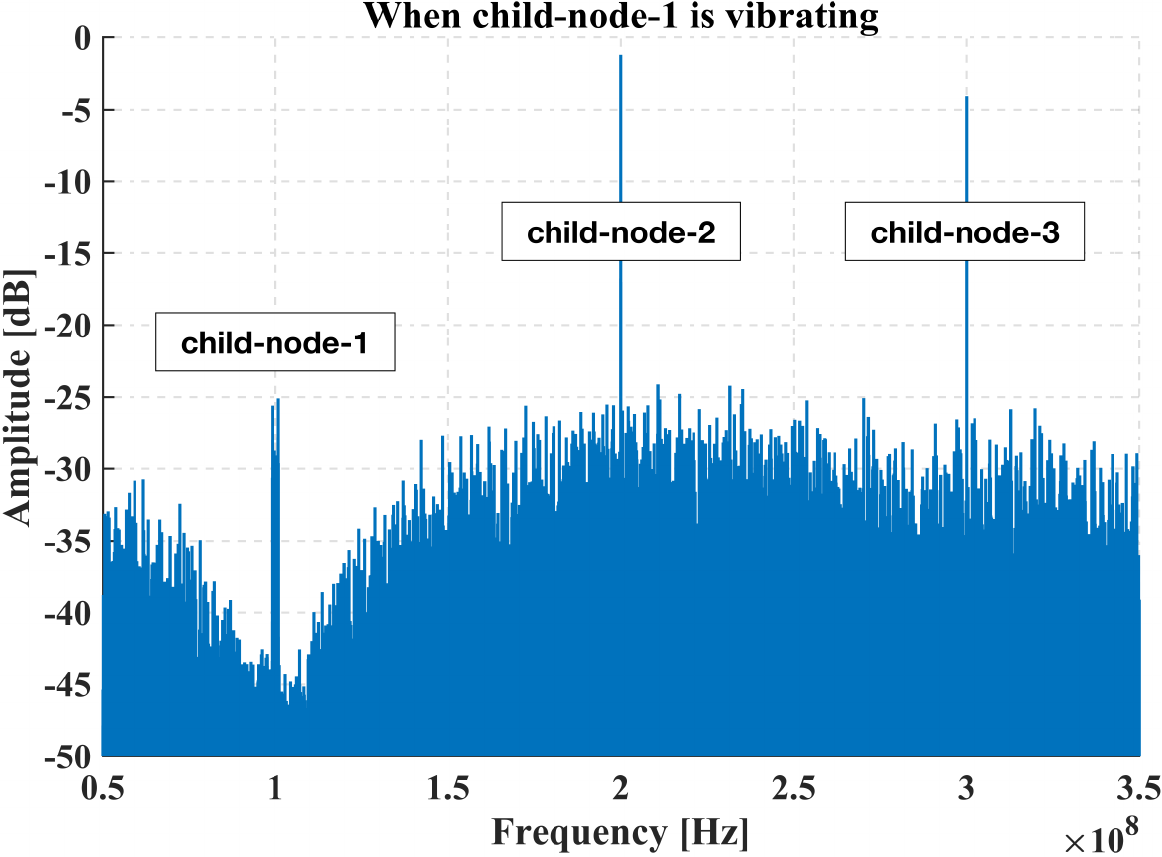}}
\hspace{3ex}
\subfigure[]{\label{fig12b}\includegraphics[width=0.3\linewidth]{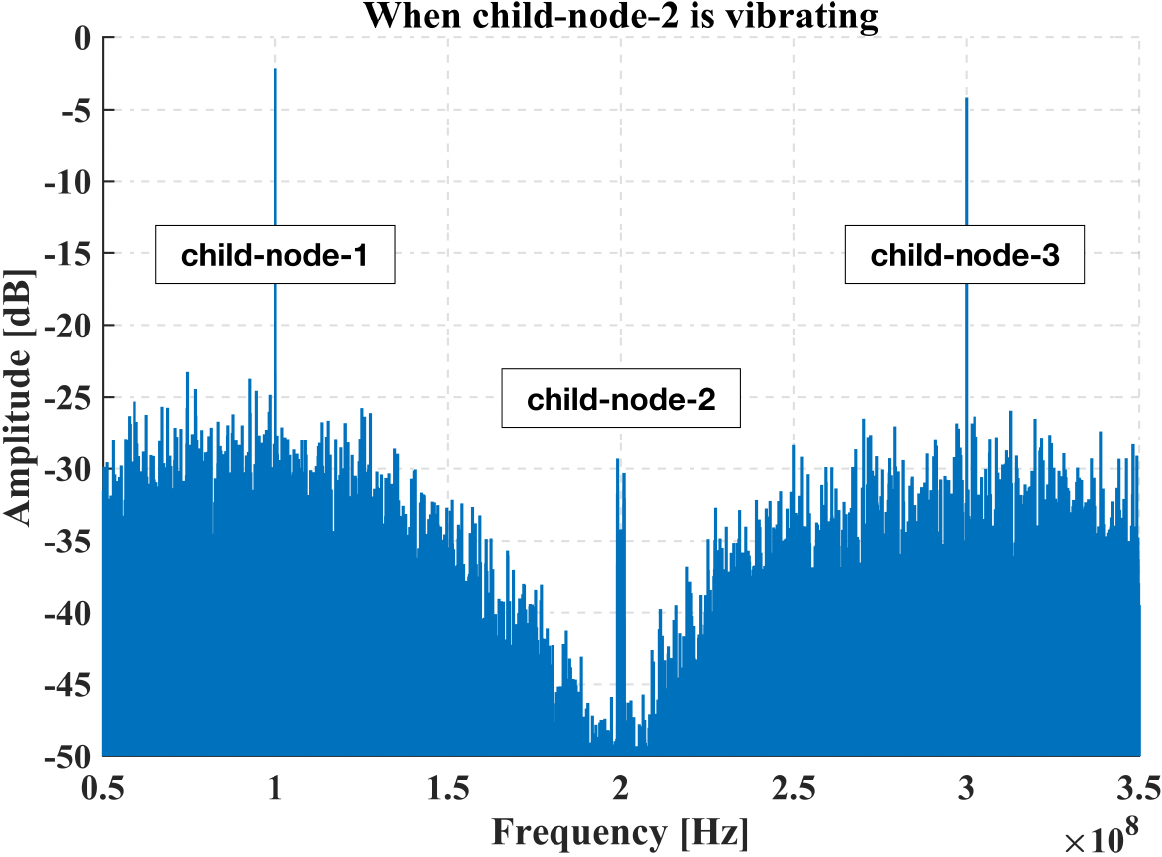}}
\hspace{3ex}
\subfigure[]{\label{fig12c}\includegraphics[width=0.3\linewidth]{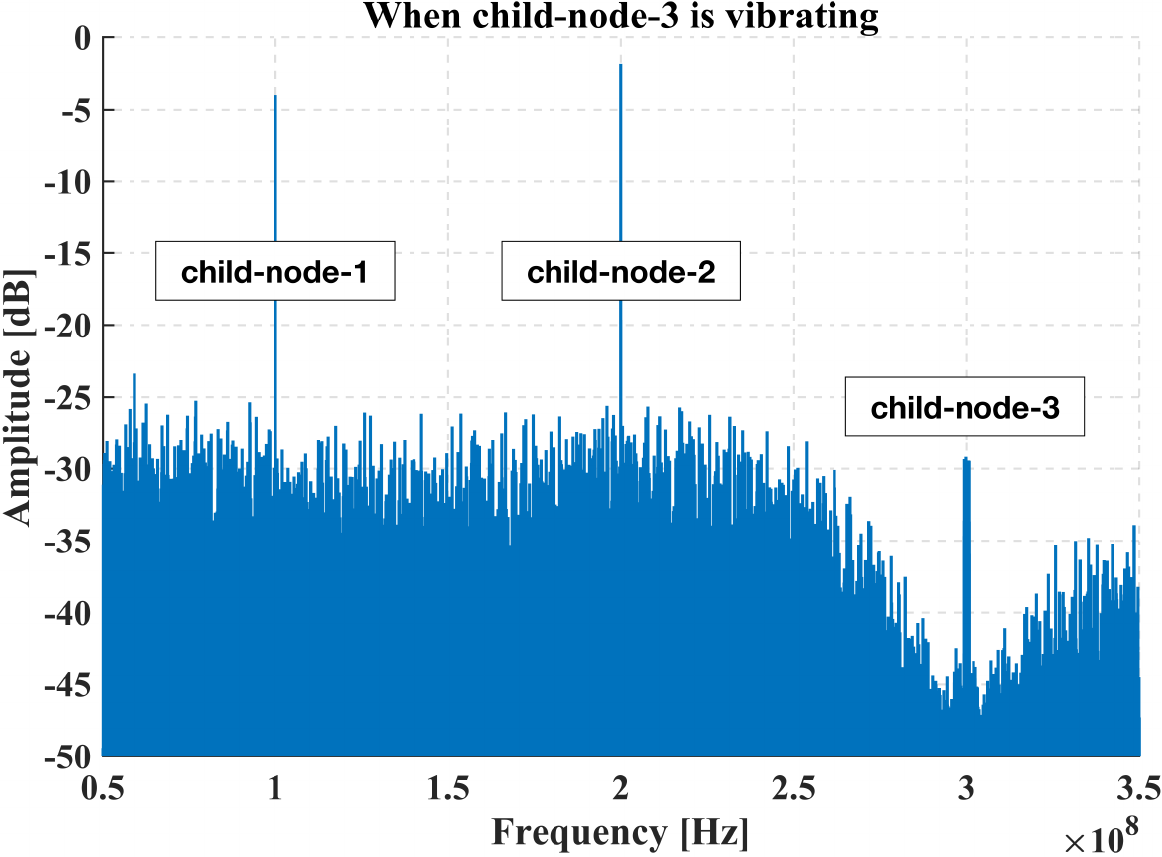}}
\caption{Frequency spectrum of the signal received by the center node when a child node is vibrating. In the (a), (b), and (c) scenarios, one child node is vibrating while the other two are not. The vibrating child node's registered band exhibits spectrum castdown and spectrum splitting.}
\label{fig12}
\end{figure*}

\begin{figure}[htb]
\centering
\includegraphics[width=1\linewidth]{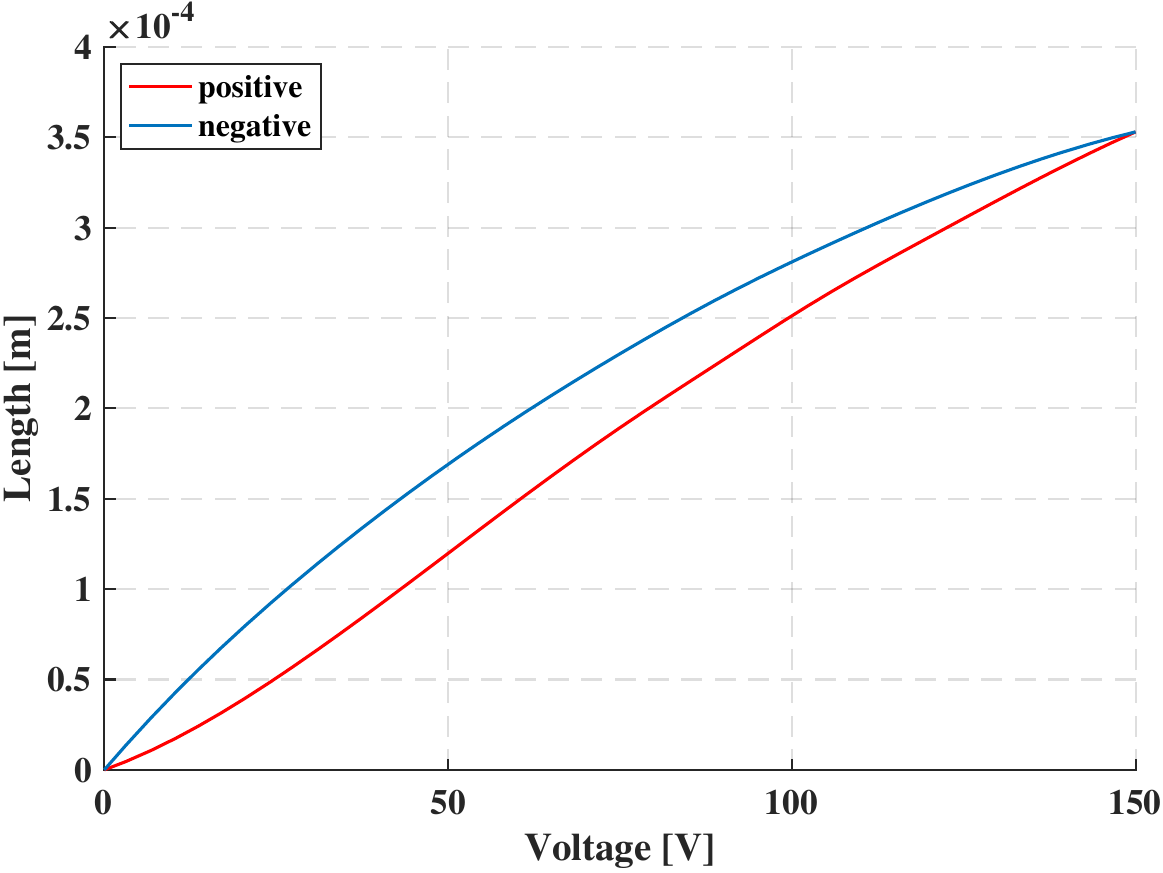}
\caption{Relationship between voltage change and length change. Due to the hysteresis effect in the open-loop PZT, the voltage change and length change are not proportional.}
\label{fig13}
\end{figure}

\begin{figure}[htb]
\centering
\includegraphics[width=1\linewidth]{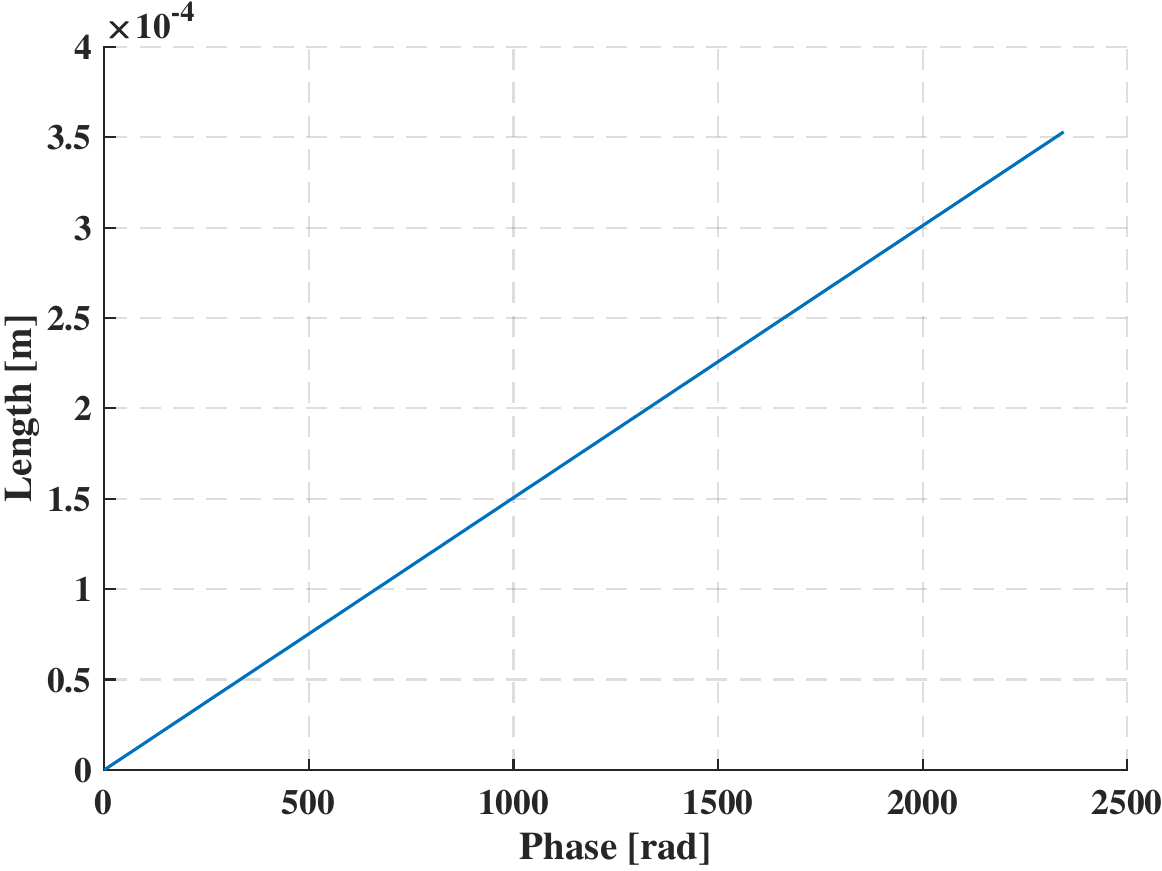}
\caption{Relationship between phase change and length change. The phase change and length change are proportional, which is consistent with the theory.}
\label{fig14}
\end{figure}

For DOFS, each child node has a PZT to simulate the vibration waveform when it reaches the child node. According to the d-type piezoelectric equation of PZT, the length change of the optical fiber wound on the PZT due to the applied radial voltage can be obtained by
\begin{equation}
\begin{aligned}
\Delta L=d \pi r \Delta E,
\end{aligned}
\end{equation}
where $d$ is the piezoelectric parameter, $r$ denotes the outer diameter, and $\Delta E$ represents the change in electric field intensity. By utilizing the relationship between electric field intensity and electric potential $E = V / t$, we can obtain
\begin{equation}
\begin{aligned}
\Delta L=\frac{d \pi r}{t} \Delta V,
\end{aligned}
\end{equation}
where $t$ is the thickness of the tube-type PZT and $\Delta V$ denotes the change in voltage. According to Eq. \ref{formula11}, the phase change of the $i$-th child node can be obtained by
\begin{equation}
\begin{aligned}
\Delta \varphi_i=\Delta V_i \frac{2 \pi}{\lambda}\left\{n-\frac{1}{2} n^3\left[(1-\mu) p_{12}-\mu p_{11}\right]\right\} \frac{d\pi r}{t},
\end{aligned}
\end{equation}
where $\Delta V_i$ represents the voltage change of PZT in the $i$-th child node. It can be observed that the voltage change $\Delta V_i$ is directly proportional to the phase change $\Delta \varphi_i$. In the experiment, Fig. \ref{fig11} displays the relationship between the phase obtained by phase unwrapping and the voltage loaded on PZT over time. The blue solid line represents the phase waveform, while the red dashed line represents the voltage waveform at the same time. This corresponds precisely to the theoretical results. Additionally, Fig. \ref{fig12} shows the frequency spectrum obtained when different child nodes sense vibration in the experiment. By observing the phenomenon of spectrum castdown or spectrum splitting, we can easily determine which child node perceives the vibration. To establish the relationship between the length change and voltage change, we conducted measurements using a capacitance micrometer with an accuracy of 0.2 $\rm{nm}$, as shown in Fig. \ref{fig13}. It indicates the existence of a hysteresis effect in the open-loop PZT, which leads to the fact that the length change does not exhibit an ideal proportional relationship with voltage change. Thus, we can find that the sinusoidal waveforms depicted in the experimental results are not perfect sinusoidal waveforms. However, we can establish the relationship between phase change and length change, as shown in Fig. \ref{fig14}. It can be observed that the phase change is proportional to the length change.

\begin{figure*}[htb]
\centering
\subfigure[]{\label{fig15a}\includegraphics[width=0.3\linewidth]{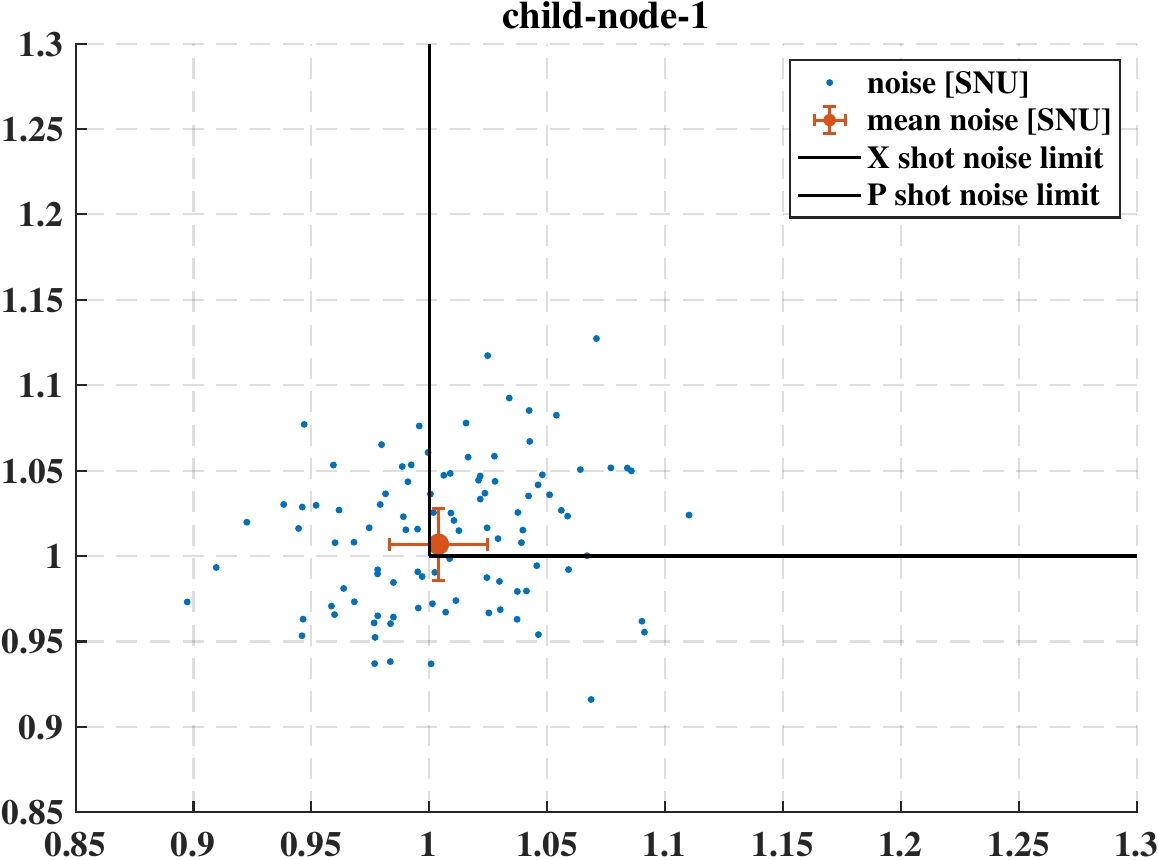}}
\hspace{3ex}
\subfigure[]{\label{fig15b}\includegraphics[width=0.3\linewidth]{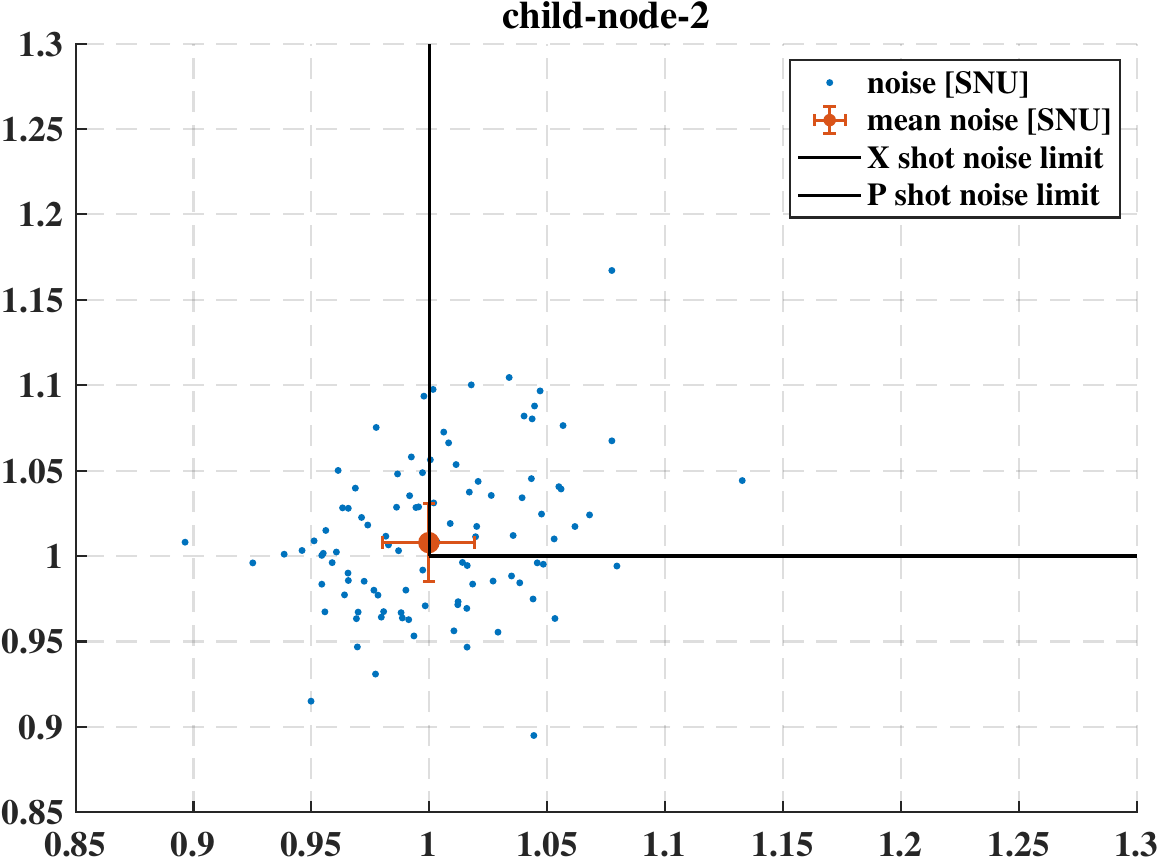}}
\hspace{3ex}
\subfigure[]{\label{fig15c}\includegraphics[width=0.3\linewidth]{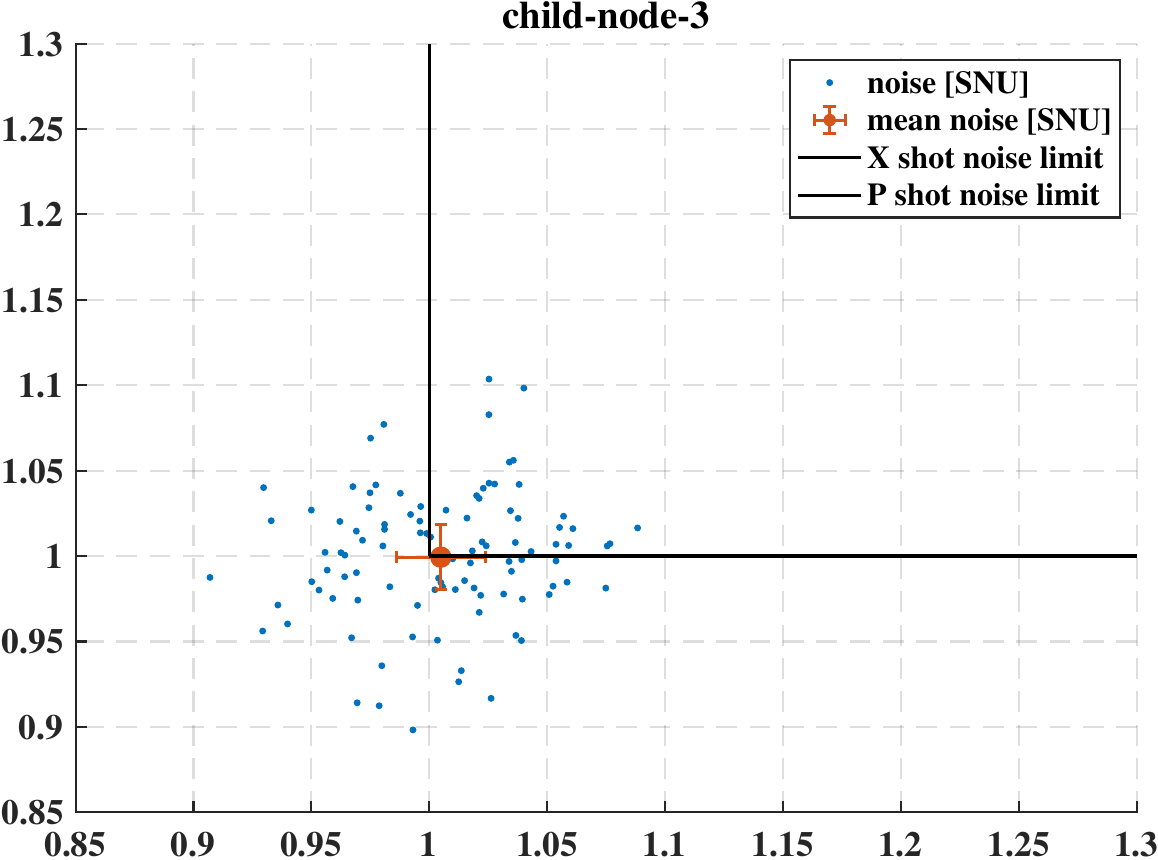}}
\caption{Detection accuracy of sensing in ISAQN. The mean detection accuracy of multiple experiments conducted by three child nodes is $(1.0038606, 1.0068994)$, $(0.9997877, 1.007847)$, and $(1.0048642, 0.9993358)$, respectively. These values closely approximate the shot noise limit values of $(1, 1)$.}
\label{fig15}
\end{figure*}

\begin{figure}[!h]
\centering
\includegraphics[width=1\linewidth]{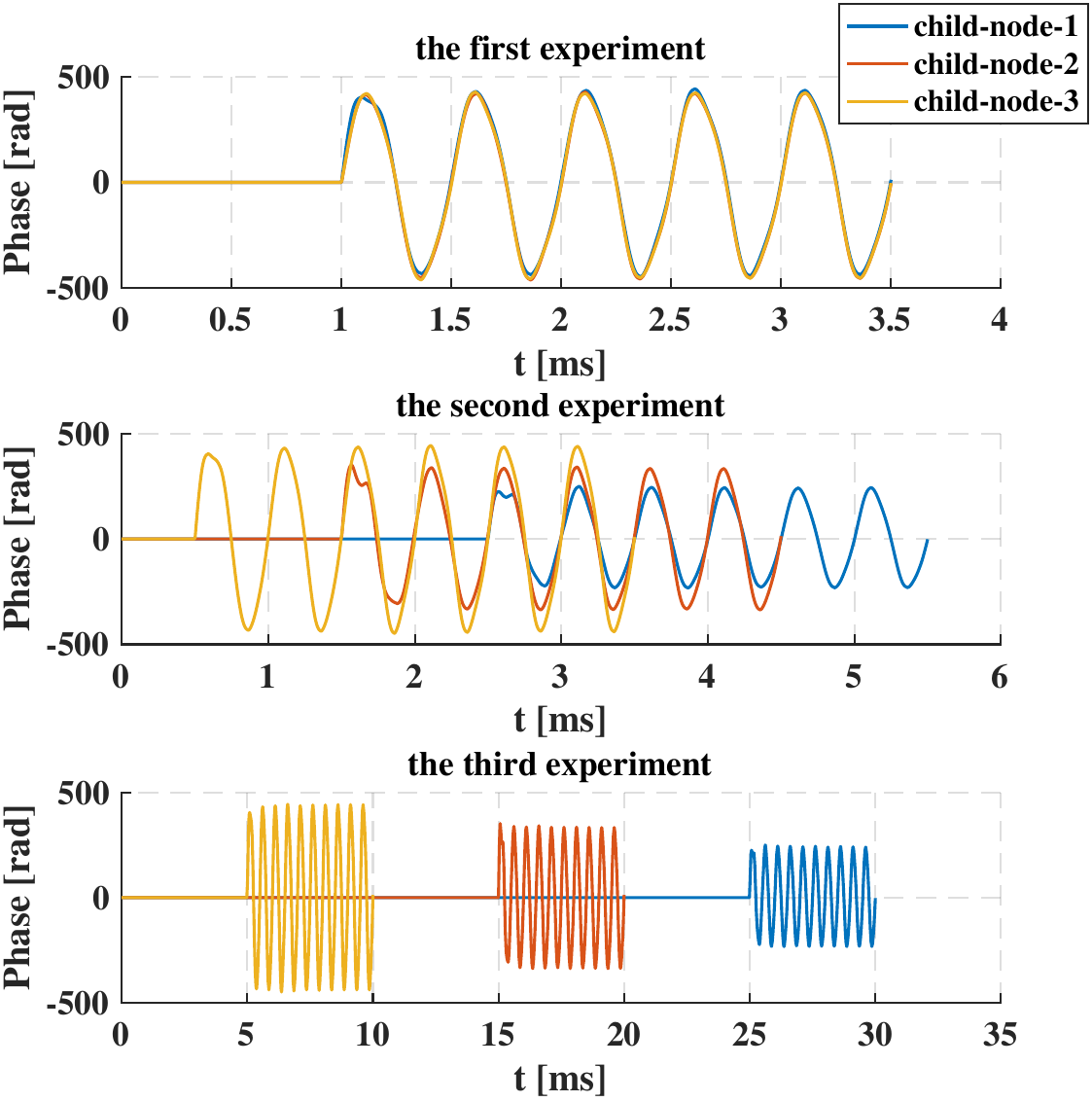}
\caption{Sensing waveforms recovered from three different scenarios. The period of the vibration waveform is 2 $\rm{kHz}$. The time differences for the start of the vibration in the three experiments are 0 $\rm{s}$, 0.001 $\rm{s}$, and 0.01 $\rm{s}$, respectively. The arrival time of the sensing waveform accurately reflects the differences in vibration start time.}
\label{fig16}
\end{figure}

The detection precision in the experiment is close to the shot noise limit, as indicated by the black line displayed in Fig. \ref{fig15}. The blue dots represent the detection precision of each individual experiment. The red dots represent the detection precision obtained by averaging over a large amount of data, with the corresponding red error bars representing the standard deviation. As shown in Fig. \ref{fig15}, the mean detection precision of multiple experiments from three child nodes are $(1.0038606, 1.0068994)$, $(0.9997877, 1.007847)$, and $(1.0048642, 0.9993358)$, which are approximately equal to the values of the shot noise limit $(1, 1)$. The spatial resolution is 0.20 $\rm{m}$, mainly limited by the bandwidth of the ICR. The maximum phase change is 891.18 $\rm{rad}$. The vibration magnitude of the vibration center can be derived according to the vibration magnitude of the child nodes. By calculating the phase power spectrum density (PSD), the noise power is around -50 $\rm{rad}^2$ $\rm{dB/Hz}$. Therefore, its strain resolution is 0.50 $\rm{n}$$\varepsilon$$/\sqrt{\rm{Hz}}$. The vibration response bandwidth ranges from 1 $\rm{Hz}$ to 2 $\rm{kHz}$.

Since the same vibration event can be detected by different nodes at different times, as shown in Fig. \ref{fig16}, we can calculate the specific location of the vibration event by the time difference. Moreover, precise time differences can be obtained using cross-correlation. In the experiment, different voltage waveforms are loaded on the PZT of three child nodes, which simulate the different effects of a vibration event on the child nodes. Assuming the velocity of the vibration event is 6 $\rm{km/s}$, three child nodes are arranged in an equilateral triangular structure, with a distance from the center node of 10 $\rm{km}$. The center node is at the center of this equilateral triangle. Under these circumstances, we can determine the center coordinate of the vibration event. Fig. \ref{fig16} shows the results of three tests. In the first vibration test, the time difference is 0 $\rm{s}$, $\Delta t_{12}=\Delta t_{23}=0$ $\rm{s}$, so the coordinates of the vibration center are located at the center node. Taking child node 1 as the reference origin, the coordinates of the vibration center are $(8660.25, 5000.00)$. In the second vibration test, the time difference is approximately 0.001 $\rm{s}$, $\Delta t_{12}=\Delta t_{23}=0.001$ $\rm{s}$. Therefore, the coordinates of the vibration center are $(8663.72, 5006.00)$. In the third vibration test, the time difference is approximately 0.01 $\rm{s}$, $\Delta t_{12}=\Delta t_{23}=0.01$ $\rm{s}$. Thus, the coordinates of the vibration center are $(8695.00, 5059.94)$. In conclusion, ISAQN has successfully achieved detection precision at the standard quantum limit, enabling the coexistence of QKD and sensing under the shot-noise-limited detection.

\section*{Discussion}

When the frequency of the vibration is too high, the pilot signal will no longer be able to totally recover the phase of the quantum signal. This is because the pilot signal is inserted into the quantum signal through the TDM method. If the pilot signal is transmitted together with the quantum signal through the FDM method, this limitation can be overcome. In this way, the recovery of the quantum signal can be achieved regardless of the vibrations at any high frequency and big amplitude. This will be our future research content. Due to space constraints, we will not continue to elaborate further.

In terms of practical security, this round-trip structure is susceptible to the eavesdropper Eve's practical security attacks, including the phase remapping attack \cite{xu2010experimental, xu2020secure} and the Trojan-Horse attack \cite{gisin2006trojan}. In order to resist the phase remapping attack, the child node can verify if the correct modulation is applied correctly by monitoring the arrival time of the reference pulse and the signal pulse \cite{xu2010experimental}. As we cannot use isolators in the round-trip structure, a filter can be used to exclude Eve's input light to prevent the Trojan-horse attack \cite{gisin2006trojan}. Moreover, three technical countermeasures exist, including the use of a watchdog mechanism with a switch at the entrance of the round-trip that randomly diverts a small fraction of incoming signals to this detector, allowing access to the eavesdropper for a shorter duration, and reducing the width of the phase modulation voltage pulse \cite{jain2014trojan}. From a theoretical perspective, a higher amount of privacy amplification can help the ISAQN to eliminate the information leakage caused by Trojan-horse attacks. It is necessary to estimate the maximum information leakage due to Trojan-horse attacks and incorporate these elements into the security proof \cite{gisin2006trojan,jain2014risk,lucamarini2015practical}.

\section*{Conclusion}

In this paper, ISAQN has been proposed and verified both theoretically and experimentally. We integrate QKD and sensing through SPM protocol. ISAQN only requires a laser and a detector to achieve point-to-multipoint QKD and DOFS. Experimental results demonstrate the network's ability to distinguish quantum signals and sensing signals from different child nodes simultaneously. This provides a new perspective for future ubiquitous quantum networks and distributed sensing.

\section*{Appendix}

Here we give the SKR calculation process for the unit system repetition rate in the asymptotic case. Firstly, the SKR for reverse reconciliation with heterodyne detection is calculated as \cite{fossier2009improvement}
\begin{equation}
K_r=\beta I_{AB}^{\rm het} - \chi_{BE}^{\rm het},
\end{equation}
where $\beta \in (0,1)$ is the efficiency of reverse reconciliation, $I_{AB}^{\rm het}$ is the mutual information between Alice and Bob, and $\chi_{BE}^{\rm het}$ is the maximum information available to Eve on Bob's key bounded by the Holevo quantity. Specifically, $I_{AB}^{\rm het}$ can be identified as
\begin{equation}
I_{AB}^{\rm het} = \log_2\frac{V+\chi_{\rm tot}}{1+\chi_{\rm tot}},
\end{equation}
where $V=V_{A}+1$, and $\chi_{\rm tot}$ representing the total noise referred to the channel input can be calculated as $\chi_{\rm tot}=\chi_{\rm line}+\chi_{\rm het}/T$, in which $\chi_{\rm line}=1/T-1+\varepsilon$, and $\chi_{\rm het}=[1+(1-\eta)+2 v_{el}]/\eta$. Besides, $\chi_{BE}^{\rm het}$ is identified as follows
\begin{equation}
\chi_{BE}^{\rm het} = \sum_{m=1}^2G\left(\frac{\lambda_m-1}{2}\right)-\sum_{m=3}^5G\left(\frac{\lambda_m-1}{2}\right),
\end{equation}
where $G(x) = (x+1)\log_2(x+1)-x\log_2x$. $\lambda_m$ are symplectic eigenvalues derived from the covariance matrices and can be expressed as
\begin{equation}
\begin{aligned}
\lambda_{1,2}^2&=\frac{1}{2}\left(A\pm\sqrt{A^2-4B}\right),\\
\lambda_{3,4}^2&=\frac{1}{2}\left(C\pm\sqrt{C^2-4D}\right),\\
\lambda_5&=1,
\end{aligned}
\end{equation}
where
\begin{equation}
\begin{aligned}
A&=V^2(1-2T) + 2T + T^2 (V+\chi_{\rm line})^2,\\
B&=T^2 (V \chi_{\rm line} +1)^2,\\
C&=\frac{1}{(T(V+\chi_{\rm tot}))^{2}}[A \chi_{\rm het}^{2}+B+1\\
&+2\chi_{\rm het}(V\sqrt{B} + T(V+\chi_{\rm line})) + 2T(V^{2}-1)],\\
D&=\left(\frac{V+\sqrt{B}\chi_{\rm het}}{T(V+\chi_{\rm tot})}\right)^{2}.
\end{aligned}
\end{equation}

\bibliography{ISAQN}

\end{document}